\documentclass[aps,twocolumn,tightenlines,nofootinbib,superscriptaddress,floatfix]{revtex4-2}
\usepackage[utf8]{inputenc}
\usepackage{graphicx}
\usepackage[caption=false]{subfig}
\usepackage{amsfonts}
\usepackage{amsmath,amssymb,bm}
\usepackage[colorlinks=true,linkcolor=blue]{hyperref}
\usepackage{booktabs}
\begin{document}
\title{Azimuthal angular correlation of \texorpdfstring{$J/\psi$}{\textit{J/ψ}} plus jet production at the electron-ion collider}

\author{Luca Maxia}
\email{l.maxia@rug.nl}
\affiliation{Van Swinderen Institute for Particle Physics and Gravity, University of Groningen, Nijenborgh 4, 9747 AG Groningen, The Netherlands}

\author{Feng Yuan}
\email{fyuan@lbl.gov}
\affiliation{Nuclear Science Division, Lawrence Berkeley National Laboratory, Berkeley, California 94720, USA}

\begin{abstract}
By investigating the soft gluon radiation in the $J/\psi$ plus jet photoproduction at the electron-ion collider (EIC), we demonstrate that the azimuthal angular correlations between the leading jet and heavy quarkonium provide a unique probe to the production mechanism of the latter.
In particular, a significant $\cos(\phi)$ asymmetry is found for the color-singlet channel, whereas it vanishes or has an opposite sign for color-octet production, depending on the jet transverse momentum. 
Numerical results of $\cos(\phi)$ and $\cos(2\phi)$ asymmetries employing both the color-singlet model and the nonrelativistic QCD approach are presented for typical kinematics at the future EIC.
\end{abstract}
\maketitle

\section{Introduction}
In recent years, heavy quarkonium production in various inclusive processes has attracted great interest as a way to probe gluon distributions both in initial (nucleon tomography) and final (fragmentation functions) states
~\cite{Godbole:2012bx, Boer:2012bt, Godbole:2013bca, Dunnen:2014eta, Mukherjee:2015smo, Mukherjee:2016cjw, Mukherjee:2016qxa, Rajesh:2018qks, Scarpa:2019fol, DAlesio:2019qpk, Kishore:2021vsm, Kishore:2022ddb, Boer:2023zit, Copeland:2023wbu, Echevarria:2023dme, Copeland:2023qed, Kang:2023doo}. 
Among them, Refs.~\cite{DAlesio:2019qpk, Kishore:2022ddb} have studied the azimuthal angular correlation in semi-inclusive DIS between $J/\psi$ and leading jet to probe the so-called linearly polarized gluon distribution. 
In this paper, we will investigate the dominant contributions from the soft gluon radiations and demonstrate that azimuthal correlations can also provide a unique opportunity to disentangle between the color-singlet (CS) and color-octet (CO) mechanisms.

\begin{figure}[t]
\begin{center}
\includegraphics[width = .7\linewidth]{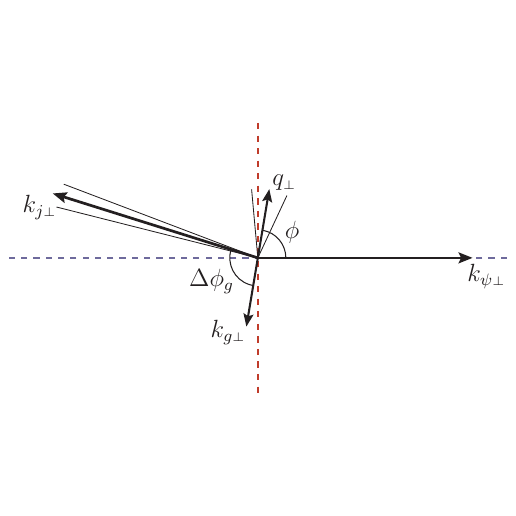}
\end{center}
\vskip -1.4cm \caption{\it Kinematic correlation between the leading jet and heavy quarkonium as viewed in the transverse plane. Here $q_\perp$ (the total outgoing transverse momentum) is small compared to individual transverse momenta.} 
\label{fig: correlation limit}
\end{figure}

In the nonrelativistic QCD (NRQCD)~\cite{Bodwin:1994jh} approach, the heavy-quark pair forms a Fock state specified by $n={}^{2S+1} L_J^{(c)}$, with $S$ denoting its spin, $L$ the orbital angular momentum, $J$ the total angular momentum and $c$ its color. Note that, within this framework, the pair can couple either as a CS or CO state. 
Therefore, comprehending the significance of the CS and CO contributions is crucial.
Although great progress has been made in understanding heavy quarkonium production in hadronic collisions (for recent reviews see~\cite{Brambilla:2010cs,Lansberg:2019adr}), challenges remain to describe quarkonium formation.
For instance, there is no formal proof of the validity of the NRQCD approach in the small transverse momentum region, and only recently works that address this issue begun to emerge~\cite{Fleming:2019pzj, Echevarria:2019ynx, Boer:2023zit, Echevarria:2024idp}. Moreover, in \cite{Flore:2020jau} it has been shown that including the next-order contributions of the CS channel greatly improves the agreement of the theoretical prediction with HERA data~\cite{H1:2010udv}, which might be an indication that the CO contributions are overestimated.
In the literature (see, e.g.,~\cite{Godbole:2012bx, Mukherjee:2015smo,Rajesh:2018qks, Mukherjee:2016cjw, DAlesio:2019qpk, Bacchetta:2018ivt, Boer:2021ehu, DAlesio:2023qyo}) it has already been vastly discussed the opportunities of future experiments at the electron-ion collider (EIC) to provide additional information on the production mechanism through cross section and/or polarization measurements.
Here, we propose an original and innovative approach to disentangle the CS and CO mechanisms at the EIC and test the significance of CO contributions (at least in the small transverse momentum region).
More specifically, we will demonstrate how azimuthal angular correlations in $J/\psi$ plus jet\footnote{Reconstruction of the jet in the final state can be achieved by applying the anti-$k_{\scriptscriptstyle T}$ algorithm.} photoproduction at the EIC offer a unique probe of the underlying production mechanism.
We will focus on the correlation limit, i.e., the transverse momentum of individual particles is much larger than the total transverse momentum. 
Therefore, by combining the transverse momenta of the $J/\psi$ ($k_{\psi\perp}$) and the jet ($k_{j\perp}$), we can identify two scales. The first one is given by $\vec{P}_\perp = \frac{\vec{k}_{\psi\perp}-\vec{k}_{j\perp}}{2}$, while the second by $\vec{q}_\perp = \vec{k}_{\psi\perp} + \vec{k}_{j\perp}$, with $| \vec q_\perp | \ll |\vec P_\perp|$.
Hence, according to this limit, the heavy quarkonium and jet are mainly produced back-to-back in the transverse plane (see Fig.~\ref{fig: correlation limit}).
An imbalance between the two final-state particles with nonzero $|\vec q_\perp|$ can be generated by high-order perturbative corrections and from the intrinsic transverse momentum of the incoming parton.
We identify this imbalance with the angle $\phi$, namely the difference between the azimuthal angles of $\vec q_\perp$ and $\vec k_{\psi \perp}$, where we can approximate the latter as $\vec k_{\psi \perp} \approx \vec P_\perp$ within the correlation limit.

Moreover, we remark that in this limit such azimuthal imbalance is mostly generated from the soft/collinear gluon radiations from perturbative diagrams (see for instance \cite{Hatta:2020bgy,Hatta:2021jcd}). This contribution, denoted by $k_{g\perp}$ in Fig.~\ref{fig: correlation limit}, tends to align with the jet direction at low $q_\perp$, which leads to significant $\cos(n\phi)$ asymmetries. 
Detailed examples have been shown for vector boson (photon/$Z$/Higgs) plus jet production in $pp$ collisions \cite{Catani:2014qha,Catani:2017tuc} and for lepton plus jet  \cite{Hatta:2020bgy} and dijet \cite{Hatta:2021jcd} productions in $ep$ collisions.
In the case of quarkonium productions, azimuthal distributions like
the $\cos(\phi)$ and $\cos(2\phi)$ can also be exploited to unravel the production mechanism. Such findings can then be applied to electroproduction to investigate gluon distributions (e.g.,~linearly polarized gluons) in nucleons and nuclei.

The remainder of the paper is the following. In Sec.~\ref{sec: one loop} we present the one loop fixed-order calculation. In Sec.~\ref{sec: CS calc} we report the derivation in the CS channel, whereas the CO is given in Sec.~\ref{sec: CO calc}. For the latter, we explicitly consider gluon and quark contributions, and discuss the importance of the LDME evolution.
In Sec.~\ref{sec: resummation and predictions} we present the resummed cross section at one loop and give numerical predictions at fixed kinematics for EIC. Conclusions are drawn in Sec.~\ref{sec: conclusions}.
In addition, our paper includes the appendix~\ref{app: CS vs CO}, where we present the resummed asymmetries and the normalized cross section in the CS and CO channels separately.

\section{Soft gluon radiation at one loop}
\label{sec: one loop}
\begin{figure}[t]
\begin{center}
\subfloat[\label{fig: SE1}]{\includegraphics[width=.31\linewidth, keepaspectratio]{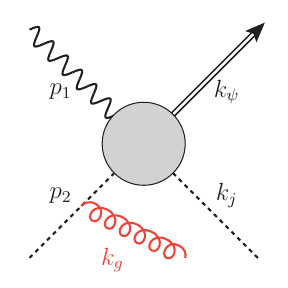}}\hfill
\subfloat[\label{fig: SE2}]{\includegraphics[width=.31\linewidth, keepaspectratio]{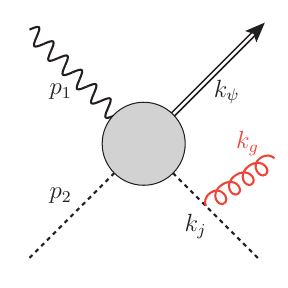}}\hfill
\subfloat[\label{fig: SE3}]{\includegraphics[width=.31\linewidth, keepaspectratio]{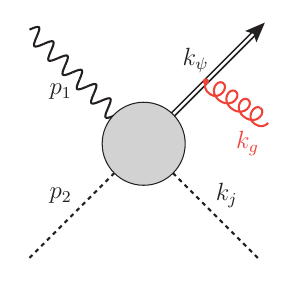}}
\end{center}
\vskip -0.4cm \caption{\it Soft gluon radiation in the $J/\psi$ plus jet photoproduction process from: (a) incoming parton, (b) outgoing parton, (c) outgoing $J/\psi$.
The parton, either a quark, antiquark or gluon, is given as a dashed line in the figure. All three diagrams contribute to the CO channel, whereas only the first two, (a) and (b), are relevant for the CS one.} 
\label{fig: soft emission}
\end{figure}

In this section we discuss the implication of azimuthal correlation between the $J/\psi$ and jet for photoproduction at the future EIC, $\gamma p\to J/\psi + jet + X$. 
The leading-order (LO) NRQCD contribution from the partonic process is given by
\begin{equation}
    \gamma(p_1) + a(p_2)\to [Q\bar Q]^{(c)}(k_\psi) + a(k_j) \ ,
\label{eq: partonic process}
\end{equation}
with $a = g$ when the heavy-quark pair $Q\bar Q$ is produced in a CS ($c=1$) configuration and ${a = g,\, q,\, \bar q}$ in the CO ($c=8$) one. Moreover, we have indicated the momentum of each particle in parentheses.
At this order, the $J/\psi$ and jet are back-to-back in the transverse plane, so that $q_\perp = 0$. 
However, at higher orders, small nonzero $q_\perp$ originates from parton intrinsic transverse momenta and soft gluon radiation.
In the following, we will derive the LO soft gluon radiation contribution (Fig.~\ref{fig: soft emission}) and the associated azimuthal angular asymmetries, whereas the collinear gluon radiation factorizes into the TMD gluon distributions. 

The major difference between CS and CO channels is that the soft gluon radiation associated with the heavy quark pair only contributes to the latter. This occurs due to cancellations between the emissions from the heavy quark and antiquark when the pair is in a CS state. This difference has significant implications for the azimuthal asymmetries, as we will discuss in the following.

We first present the results of the soft gluon radiation at fixed order, while we will present the resummed form in Sec.~\ref{sec: resummation and predictions}.

\subsection{Color-singlet channel}
\label{sec: CS calc}

Starting with the CS channel, by adding Fig.~\ref{fig: SE1} and Fig.~\ref{fig: SE2} we obtain the amplitude squared (averaged over the color and spin of incoming particles) for the soft gluon radiation:
\begin{equation}
    |\overline{{\cal A}_1^{(1)}}|^2 = g_s^2 C_A\, |\overline{{\cal A}_0^{g, (1)}}|^2\, S_g(p_2, k_j)\ ,
\label{eq: A1 CS}
\end{equation}
where $A_0^{g, (1)}$ is the LO amplitude and $S_g(v_a, v_b)$ is a shorthand notation for
\begin{equation}
    S_g(v_a, v_b)=\frac{2\, (v_a\cdot v_b)}{(v_a\cdot k_g) (v_b\cdot k_g)} \ .
\label{eq: Sg definition}
\end{equation}
More specifically,
\begin{align}
    S_g(p_2, k_j)& = 
    \frac{2\, (p_2\cdot k_j)}{(p_2 \cdot k_g) (k_j \cdot k_g)} \nonumber\\
    &=
    \frac{2}{| \vec k_{g\perp}|^2} \frac{e^{\Delta y_{g}}}{\cosh(\Delta y_{g}) - \cos(\Delta \phi_{g})}\nonumber\\
    & \approx \frac{2}{| \vec k_{g\perp}|^2} \left(1 + \frac{\sinh(\Delta y_g)}{\cosh(\Delta y_g) -\cos(\phi)}\right.\nonumber\\
    &\left.~~~+ \frac{\cos(\phi)}{\cosh(\Delta y_g) - \cos(\phi)}\right)\ ,
\label{eq: Sg(p2,kj) splitting}
\end{align}
{where  we have already separated $S_g(p_2, k_2)$ into three contributions for convenience later. We have defined $\Delta y_g = y_g - y_j$, namely the difference between the emitted soft gluon and jet rapidities, while the approximation sign is due to the relation between $\Delta \phi_g$ and $\phi$, ${\Delta \phi_g \approx \phi}$.
We need to integrate over the phase space of the emitted soft gluon to derive the leading contribution, 
\begin{align}
    & \int \frac{{\rm d}^3k_g}{(2\pi)^3 2 E_{k_g}}\, 
    |\overline{{\cal A}_1^{(1)}}|^2\, \delta^{(2)} (q_\perp+k_{g\perp})\nonumber \\ 
    &~~ = \frac{\alpha_s C_A}{2\pi^2 |\vec q_{\perp}|^2}|\overline{{\cal A}_0^{(1)}}|^2\left[\ln\frac{\hat s}{| \vec q_{\perp}|^2}+ \ln \frac{\hat t}{\hat u} + I_j(R, \phi) \right]\  , 
\label{eq: int CS}
\end{align}
where ${\hat s = (p_1 + p_2)^2}$, ${\hat t = (p_2 - k_j)^2}$ and ${\hat u = (p_1 - k_j)^2}$.
The first term in the bracket of Eq.~\eqref{eq: Sg(p2,kj) splitting} leads to the double-logarithm. The second one, being an odd function of $\Delta y_g$, receives contributions only from the boundaries of the integration region, which causes the presence of the additional logarithmic term, $\ln({\hat t}/{\hat u})$. 
The last term in Eq.~\eqref{eq: Sg(p2,kj) splitting} contains the jet contribution to azimuthal angular asymmetries.

Although its integral was relevant, and therefore computed, in other works~\cite{Sun:2014gfa,Sun:2015doa,Liu:2018trl,Liu:2020dct,Hatta:2020bgy,Hatta:2021jcd}, for completeness we report the calculation in the following.
To better analyze the physical content of this derivation, we can divide $S_g(p_2, k_2)$ into three contributions as follows
\begin{align}
    S_g(p_2, k_j)& = 
    \frac{2\, (p_2\cdot k_j)}{(p_2 \cdot k_g) (k_j \cdot k_g)} \nonumber\\
    &=
    \frac{2}{| \vec k_{g\perp}|^2} \frac{e^{\Delta y_{g}}}{\cosh(\Delta y_{g}) - \cos(\Delta \phi_{g})}\nonumber\\
    & \approx \frac{2}{| \vec k_{g\perp}|^2} \left(1 + \frac{\sinh(\Delta y_g)}{\cosh(\Delta y_g) -\cos(\phi)}\right.\nonumber\\
    &\left.~~~+ \frac{\cos(\phi)}{\cosh(\Delta y_g) - \cos(\phi)}\right)\ ,
\label{eq: Sg(p2,kj) splitting}
\end{align}
where we have defined $\Delta y_g = y_g - y_j$, namely the difference between the emitted soft gluon and jet rapidities.
The first term in the bracket of Eq.~\eqref{eq: Sg(p2,kj) splitting} leads to the double-logarithm (in $b_{\scriptscriptstyle T}$ space). The second one, being an odd function of $\Delta y_g$, receives contributions only from the boundaries of the integration region, causing the presence of the additional logarithmic term, $\ln({\hat t}/{\hat u})$, which depends on the jet rapidity.
The last term in Eq.~\eqref{eq: Sg(p2,kj) splitting} contains the jet contribution to azimuthal angular asymmetries.
The third term, $I_j$, is one of the subjects of this work, being azimuthal distribution that arises from the soft gluon radiation. 
As a result of the removal of collinear divergences already included within the jet function, $I_j$ depends on the jet size $R$.

In particular,
\begin{align}
    I_j(R,\phi) & = \int {\rm d} \Delta y_g\, \frac{\cos(\phi)}{\cosh(\Delta y_g) - \cos(\phi)} \nonumber\\
    &\quad- \frac{| \vec k_{g\perp}|^2}{2}  \int {\rm d} \Delta y_g\, S_g(p_2, k_j)\, \Theta (\Delta_{k_g k_j} < R^2) \nonumber \\
    & = \int {\rm d} \Delta y_g\, \frac{\cos(\phi)}{\cosh(\Delta y_g) - \cos(\phi)} \Theta (\Delta_{k_jk_g}>R^2) \nonumber\\
    &\quad- 2\sqrt{R^2 + \phi^2} \ ,
\label{eq: Ij definition}
\end{align}
where $\Theta (\Delta_{k_jk_g}\lessgtr R^2)$ implies that the integration is restricted inside ($<$) or outside ($>$) the rapidity region occupied by the jet cone, namely
\begin{equation}
    |\Delta y_g| > \sqrt{R^2 + \phi^2}\ .
\end{equation}
To further investigate this distribution, we expand it in a Fourier series according to
\begin{equation}
    I_j (R,\phi) = C_0^{(j)} (R) +  2 \sum_{n=1}^\infty C_n^{(j)} (R)\, \cos(n \phi)\ .
\label{eq: Fourier Series of I_j}
\end{equation} 
For a general $R$, the Fourier expansion of $I_j(R, \phi)$ is manageable only via computational methods. However, the analytical evaluation of this expansion within the small-$R$ limit is possible, giving
\begin{align}
    I_j(R,\phi) & = \ln\frac{1}{R^2} + 2 \cos(\phi) \left(\ln\frac{1}{R^2} + 2\ln(4) - 2\right) \nonumber\\
    & \phantom{=} + 2 \cos(2\phi) \left(\ln\frac{1}{R^2} - 1 \right) + \cdots \ .
\label{eq: I_j approx Fourier series}
\end{align}

\begin{figure}[t]
\begin{center}
\includegraphics[width = .9\linewidth]{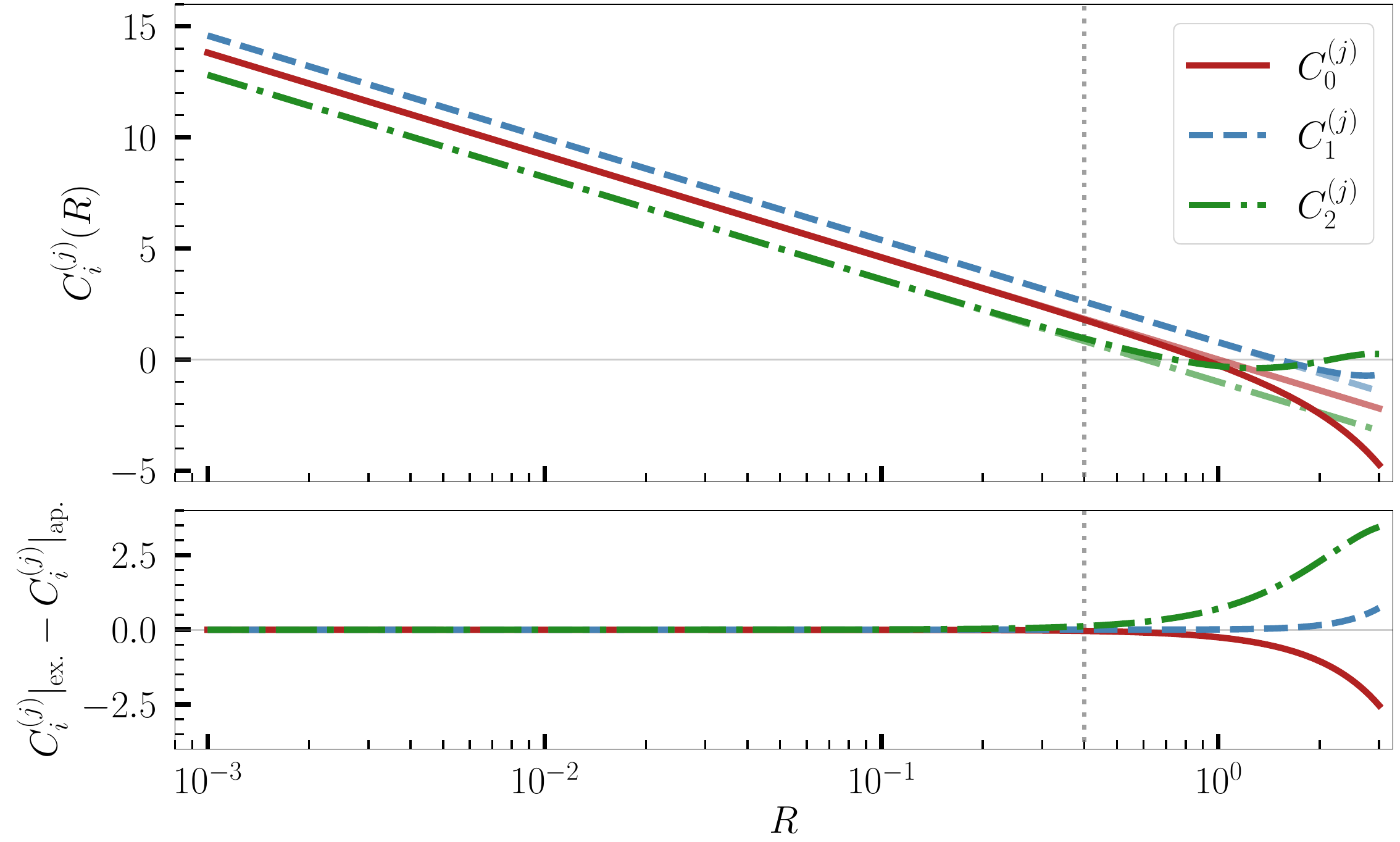}
\end{center}
\vskip -0.4cm \caption{\it Dependence of the first coefficients of Eq.~\eqref{eq: Fourier Series of I_j} with respect to $R$. In the upper panel we show both the exact result obtained from a numerical computation (full color) and the approximated one given in Eq.~\eqref{eq: I_j approx Fourier series} (softer color). In the lower panel, we present the difference between the two. The vertical dotted line corresponds to $R = 0.4$, beyond which the approximation fails.}
\label{fig: coeff jet}
\end{figure}

In Fig.~\ref{fig: coeff jet}, we show how the first three coefficients of the Fourier expansion behave with respect to $R$. From this figure, we conclude that the approximation used in Eq.~\eqref{eq: I_j approx Fourier series} is reasonably adequate for $R < 0.4$.
Moreover, in line with previous works~\cite{Sun:2014gfa,Sun:2015doa,Liu:2018trl,Liu:2020dct,Hatta:2020bgy,Hatta:2021jcd}, we find that the soft gluon radiation associated with the jet leads to a dominant $\cos(\phi)$ asymmetry.

\subsection{Color-octet channel}
\label{sec: CO calc}

At variance with the CS case, when the pair forms a CO state all diagrams in Fig.~\ref{fig: soft emission} are relevant.
While it is straightforward to evaluate the soft gluon emission from the first two diagrams and prove that it does not depend on the CO Fock state, it is worthwhile to elaborate more on the last one (Fig.~\ref{fig: SE3}), which corresponds to the soft gluon radiation from the heavy-quark pair itself.
We can separate this last contribution in two terms: one that is also independent of the CO Fock state and the other that mixes $S$- and $P$-wave due to the LDME evolution~\cite{Fleming:2019pzj, Butenschoen:2019lef, Butenschoen:2020mzi, Boer:2023zit, Echevarria:2024idp} (higher states are involved too, but suppressed according to the velocity expansion of NRQCD).
The former is similar to the result obtained for the soft gluon radiation from a gluon jet in the final state~\cite{Sun:2015doa}, with the only difference originating from the gluon on-shell condition, $k^2 = 0$ for jets and $k^2 = M_V^2 \approx 4M_Q^2$ for quarkonia.
The second contribution occurs via the emission of soft gluons of order $M v$, where $v$ is the relative velocity of the heavy-quark pair. 
Since the soft gluons considered here have momenta of order $q_\perp$ (which can be greater than $M v$), one might expect a suppression of this contribution. In the next subsection, we will show that the LDME evolution actually plays a significant role in the predictions of the asymmetries.
Note that, momentarily, we will keep these Fock-state dependent contributions implicit, referring to them as “\textit{mixing.}” More details will be provided in Sec.~\ref{sec: ldme mix}.

With this in mind, we can combine the soft gluon radiations from the initial and final state gluons (Figs.~\ref{fig: SE1} and~\ref{fig: SE2}), and the averaged CO amplitude squared is summarized as follows
\begin{align}
    |\overline{{\cal A}_1^{(8)}}|^2 & = g_s^2
    C_A\Bigg\{|\overline{{\cal A}_0^{g, (8)}}|^2\, \bigg[ S_g(p_2, k_j) + \frac{1}{2} \Big( S_g(p_2, k_\psi) \nonumber\\
    & \phantom{=} - S_g(k_\psi, k_\psi) + S_g(k_j,k_\psi) - S_g(p_2, k_j)\Big) \bigg]\nonumber\\
    & \phantom{=} + \sum_q
    |\overline{{\cal A}_0^{q, (8)}}|^2\, \bigg[\frac{C_F}{C_A}\, S_g(p_2, k_j) + \frac{1}{2} \Big( S_g(p_2, k_\psi) \nonumber\\
    & \phantom{=} - S_g(k_\psi, k_\psi) + S_g(k_j,k_\psi)  - S_g(p_2, k_j)\Big) \bigg] \Bigg\} \nonumber\\
     & \phantom{=} + \textit{mixing}\ ,
     \label{eq: A1 CO}
\end{align}
where we have taken into account the contributions from both gluon and quark channels and ${\cal A}_0^{g, (8)}$ (${\cal A}_0^{q, (8)}$) represents the LO gluon (quark) amplitude. The relative importance depends on the kinematics~\cite{Copeland:2023qed}.
Note that the first term of Eq.~\eqref{eq: A1 CO} is equivalent to that in Eq.~\eqref{eq: A1 CS} for the CS case and, therefore, is a contribution purely driven by the gluon jet, whereas the corresponding term in the quark sector differs by a Casimir scaling factor.
On the other hand, all the other terms gathered in the curved parentheses are associated with the $J/\psi$ and always proportional to $C_A$.
Now, as before, we need to carry out the integration over the phase space of the emitted soft gluon. We discuss the integral of each (new) $S_g$ term separately, while the overall contribution is given at the end of this section.

\textit{\textbf{Integral of $\bm{S_g(p_2, k_\psi)}$.}}
Similarly to $S_g(p_2, k_j)$, we divide $S_g(p_2, k_\psi)$ in the following three terms
\begin{align}
    &S_g(p_2, k_\psi) = 
    \frac{2\, (p_2 \cdot k_\psi)}{(p_2 \cdot k_g) (k_\psi \cdot k_g)} \nonumber\\
    &~~=
    \frac{2}{| \vec k_{g\perp}|^2}  \frac{\sqrt{1 + m_{\psi\perp}^2}\, e^{\Delta y_{g\psi}}}{\sqrt{1 + m_{\psi\perp}^2} \cosh(\Delta y_{g\psi}) - \cos(\Delta \phi_{g\psi})}\nonumber\\
    &~~ \approx \frac{2}{| \vec k_{g\perp}|^2} \left(1 + \frac{\sqrt{1 + m_{\psi\perp}^2}\,  \sinh(\Delta y_{g\psi})}{\sqrt{1 + m_{\psi\perp}^2}\, \cosh(\Delta y_{g\psi}) + \cos(\phi)} \right.\nonumber\\
    &\left.~~- \frac{\cos(\phi)}{\sqrt{1 + m_{\psi\perp}^2}\, \cosh(\Delta y_{g\psi}) + \cos(\phi)}\right)\ ,
\label{eq: Sg(p2,kpsi) splitting}
\end{align}
where we have defined $\Delta y_{g\psi} = y_g - y_\psi$, namely the difference between the emitted soft gluon and $J/\psi$ rapidities, and $m_{\psi\perp} = M_\psi/|\vec P_\perp|$, with $M_\psi$ being the $J/\psi$ mass. Moreover $\Delta \phi_{g\psi} = \phi_g - \phi_\psi = \phi_g$  in a frame where $\phi_\psi = 0$. 
As for Eq.~\eqref{eq: Sg(p2,kj) splitting}, the first term in the bracket of Eq.~\eqref{eq: Sg(p2,kpsi) splitting} leads to the double-logarithm while the second one receives contributions only from the boundaries of the integration region, providing the logarithms: ${\ln \frac{\hat u}{\hat t} + \ln \frac{1 - M_\psi^2/\hat u}{1 - M_\psi^2/\hat t}}$.
These are once again dependent on the rapidities, but in this case both $y_j$ and $y_\psi$.
The last term in Eq.~\eqref{eq: Sg(p2,kpsi) splitting} contains part of the $J/\psi$ contribution to azimuthal angular asymmetries. 

\begin{figure}[t!]
\begin{center}
\includegraphics[width =.9\linewidth]{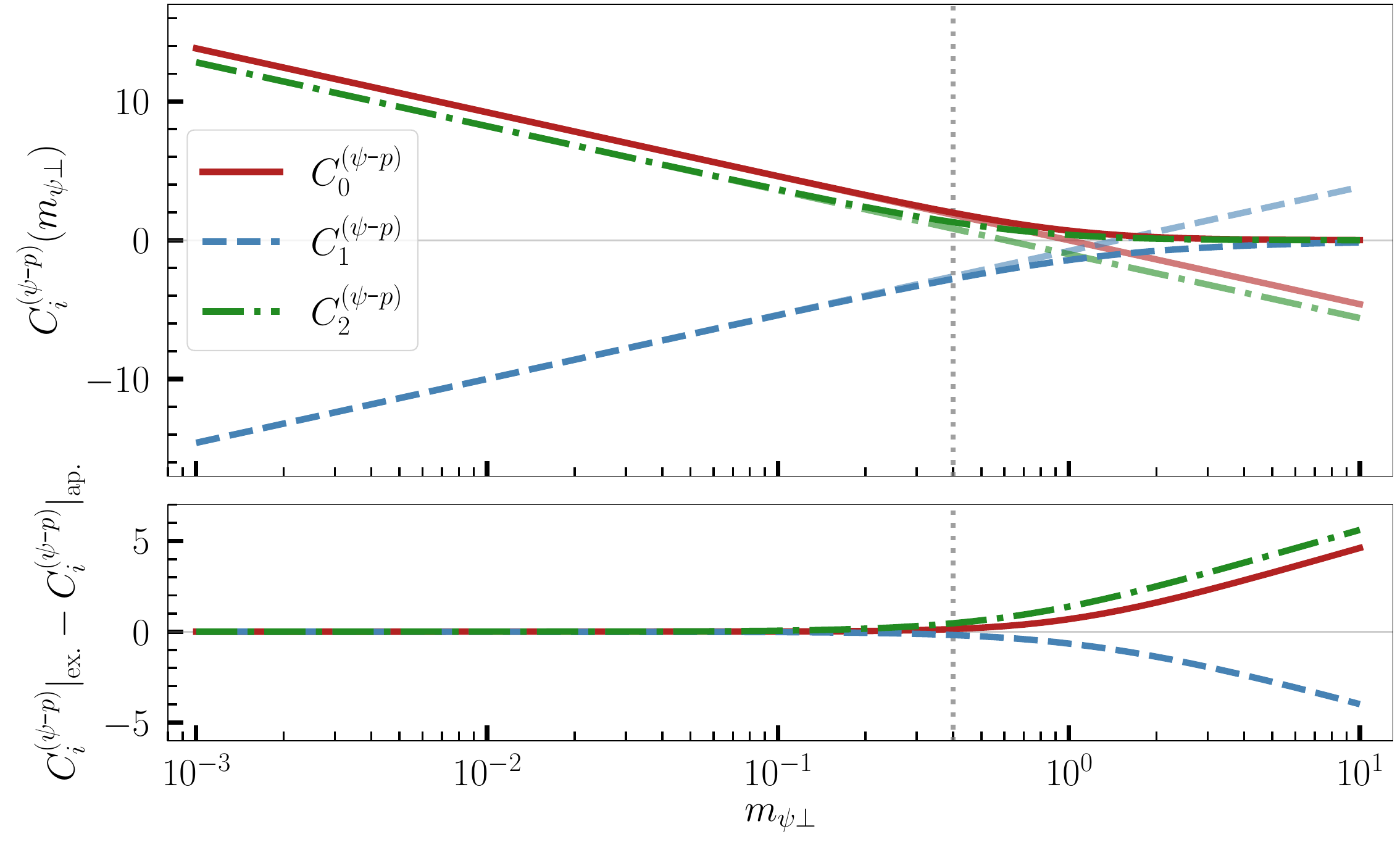}
\end{center}
\vskip -0.4cm \caption{\it Dependence of the first coefficients of Eq.~\eqref{eq: I_psi1 Fourier series} with respect to $m_{\psi\perp}$. The vertical dotted line corresponds to $m_{\psi\perp} = 0.4$. Panels follow the same logic as Fig.~\ref{fig: coeff jet}.}
\label{fig: coeff Jpsi 1}
\end{figure}

We identify this angular distribution as
\begin{align}
    &I_{\psi \text{-} p}(m_{\psi \perp}, \phi)\nonumber\\
    &\quad = \int {\rm d} \Delta y_{g\psi} \left[ -\frac{\cos(\phi)}{\sqrt{1 + m_{\psi\perp}^2}\, \cosh(\Delta y_{g\psi}) + \cos(\phi)} \right] 
\label{eq: I_psi-jet}
\end{align}
Note that the presence of the mass in the denominator acts as a regulator, and therefore $I_{\psi \text{-} p}$ is continuous for all values of $\phi$.
Moreover, when evaluated within the jet rapidity region, Eq.~\eqref{eq: Sg(p2,kpsi) splitting} leads to
\begin{align}
& I_{\psi \text{-} p}^{\rm jet} (R, m_{\psi\perp}, \Delta y, \phi) \nonumber\\
&\quad= \frac{| \vec k_{g\perp}|^2}{2} \int {\rm d} \Delta y_{g}\, S_g(p_2, k_\psi)\, \Theta (\Delta_{k_g k_j} < R^2) \ ,
\label{eq: Sg(p2, kpsi) in the jet}
\end{align}   
which not only depends on both parameters $R$ and $m_{\psi\perp}$, but also on the rapidity difference $\Delta y = y_\psi - y_j$ between the $J/\psi$ ($y_\psi$) and the leading jet ($y_j$). 
The distribution in Eq.~\eqref{eq: I_psi-jet} can be expanded in the Fourier series
\begin{align}
    & I_{\psi \text{-} p}(m_{\psi\perp}, \phi) \nonumber\\
    &\quad = C_{0}^{(\psi \text{-} p)}(m_{\psi \perp}) + 2\, \sum_{n=1}^\infty C_{n}^{(\psi \text{-} p)} (m_{\psi \perp})\, \cos(n \phi)\ ,
\label{eq: I_psi1 Fourier series}
\end{align}
where the coefficients can be analytically evaluated only in the small-$m_{\psi\perp}$ limit, for which
\begin{align}
    & I_{\psi \text{-} p}(m_{\psi\perp}, \phi) \nonumber\\
    & \quad = \ln\frac{1}{m_{\psi\perp}^2} - 2 \cos(\phi) \left(\ln\frac{1}{m_{\psi\perp}^2} + 2\ln(4) - 2\right) \nonumber\\
    & \quad \phantom{=} + 2 \cos(2\phi) \left(\ln\frac{1}{m_{\psi\perp}^2} - 1 \right) + \cdots  \ . 
\label{eq: I_psi1 approx Fourier series}
\end{align}

Fig.~\ref{fig: coeff Jpsi 1} shows the dependence of the first three coefficients with respect to $m_{\psi\perp}$, together with the reliability of the approximation introduced in Eq.~\eqref{eq: I_psi1 approx Fourier series}.

\textit{\textbf{Integral of $\bm{S_g(k_j, k_\psi)}$.}}
Compared to the previous functions, deriving the azimuthal distribution arising from $S_g(k_j,k_\psi)$ requires some extra care. Firstly, we recast the function as follows
\begin{widetext}
\begin{align}
    S_g(k_j, k_\psi) & = 
    \frac{2\, (k_j \cdot k_\psi)}{(k_j \cdot k_g) (k_\psi \cdot k_g)} 
    \approx
    \frac{2}{|\vec k_{g\perp}|^2}  \frac{\sqrt{1 + m_{\psi\perp}^2}\, \cosh(\Delta y) + 1}{\Big[\cosh(\Delta y_{g}) - \cos(\phi) \Big]\Big[\sqrt{1 + m_{\psi\perp}^2} \cosh(\Delta y_{g\psi}) + \cos(\phi) \Big]}\nonumber\\ 
    & = \frac{2}{|\vec k_{g\perp}|^2} 
    \Bigg(  \frac{\cos (\phi)}{ \cosh (\Delta y_g) - \cos(\phi)}  - \frac{\cos (\phi)}{\sqrt{1 + m_{\psi\perp}^2} \cosh (\Delta y_{g\psi}) + \cos (\phi)} \nonumber\\
    & \phantom{=} \hspace{2cm} + \frac{\sinh(\Delta y_g)}{\cosh(\Delta y_g) - \cos(\phi)} + \frac{\sqrt{1 + m_{\psi\perp}^2}\,  \sinh(\Delta y_{g\psi})}{\sqrt{1 + m_{\psi\perp}^2}\, \cosh(\Delta y_{g\psi}) + \cos(\phi)} + \widehat S_g(k_j, k_\psi) \Bigg)\ ,
\label{eq: Sg(kj,kpsi) splitting}
\end{align} 
\end{widetext}
where we remark that $\Delta y = y_\psi - y_j$, $\Delta y_g = y_g - y_j$ and $\Delta y_{g\psi} = y_j - y_\psi$, with $y_j$, $y_\psi$ and $y_g$ being respectively the jet, $J/\psi$ and emitted soft gluon rapidities.
The first four terms of Eq.~\eqref{eq: Sg(kj,kpsi) splitting} coincide with the last terms of Eqs.~\eqref{eq: Sg(p2,kj) splitting} and~\eqref{eq: Sg(p2,kpsi) splitting}, respectively, and thus remove the double counting in the azimuthal dependences. 
We identify the azimuthal distribution driven by the first two terms as
\begin{align}
&    I_{\psi \text{-} j} (R, m_{\psi\perp}, \phi)  = \int {\rm d} y_g \, 
    \Bigg( \frac{\cos (\phi)}{ \cosh (\Delta y_g) - \cos(\phi)} \nonumber\\
    &~~~ - \frac{\cos (\phi)}{\sqrt{1 + m_{\psi\perp}^2} \cosh (\Delta y_{g\psi}) + \cos (\phi)} \Bigg) \ ,
\end{align}
independent of $\Delta y$.
The last term of Eq.~\eqref{eq: Sg(kj,kpsi) splitting}, which is explicitly given by
\begin{widetext}
\begin{align}
    \widehat S_g(k_j, k_\psi) 
        & = \frac{1}{\Big[ \cosh(\Delta y_g) - \cos(\phi) \Big]\, \Big[ \sqrt{1 - m_{\psi \perp}^2} \cosh(\Delta y_{g\psi}) + \cos(\phi) \Big]} \nonumber \bigg[ \Big(\cosh(\Delta y_g) - \sinh(\Delta y_g) \Big)\, \cos(\phi) - \cos(2\phi)\nonumber \\
    & \phantom{=}~~ \phantom{\times}~~ - \sqrt{1 - m_{\psi \perp}^2} \Big( \cosh(\Delta y) - \cosh(\Delta y_{g\psi}) \big( \sinh(\Delta y_g) + \cos(\phi) \big) - \sinh(\Delta y_{g\psi}) \big( \sinh(\Delta y_g) - \cos(\phi) \big) \Big) \bigg]  \ , 
\label{eq: Sg(kj,kpsi) reduced}
\end{align}
\end{widetext}
produces the unique azimuthal distribution of $S_g(k_j, k_\psi)$ 
\newpage
\begin{equation}
    I_{\psi \text{-} j} (m_{\psi\perp}, \Delta y, \phi) = \int {\rm d} y_g \, \widehat S_g(k_j, k_\psi) \ .
\label{eq: I_psi2}
\end{equation}
Moreover, $\widehat S_g(k_j, k_\psi)$ provides another azimuthal distribution when evaluated within the jet region
\begin{align}
    & I_{\psi \text{-} j}^{\rm jet} (R, m_{\psi\perp}, \Delta y, \phi) \nonumber\\
    & \quad = \frac{|\vec k_{g\perp}|^2}{2} \int {\rm d} \Delta y_g\, \widehat S_g(k_j, k_\psi)\, \Theta(\Delta_{k_j k_\psi} < R^2)\ .
\end{align}

\begin{figure}[t]
\begin{center}
\includegraphics[width = .9\linewidth]{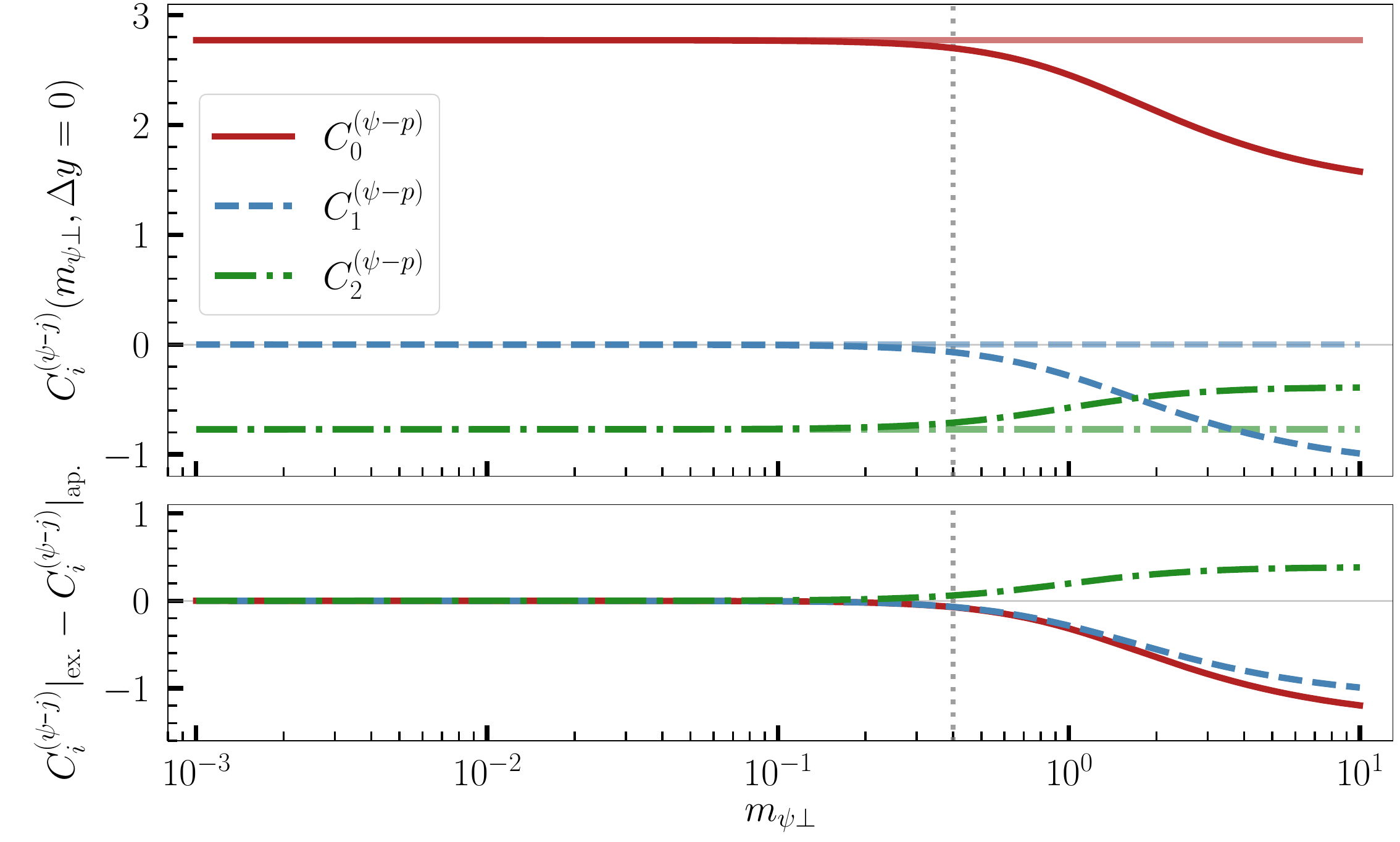}
\end{center}
\vskip -0.4cm \caption{\it Dependence of the first coefficients of Eq.~\eqref{eq: I_psi2 Fourier series} with respect to $m_{\psi\perp}$ and for $\Delta y = 0$. The coefficient $C_1$, being zero, is not shown here. Vertical dotted line corresponds to $m_{\psi\perp} = 0.4$. Panels follow the same logic as Fig.~\ref{fig: coeff jet}.}
\label{fig: coeff Jpsi 2 (Dy=0)}
\end{figure}

\begin{figure}[t]
\begin{center}
\includegraphics[width = .9\linewidth]{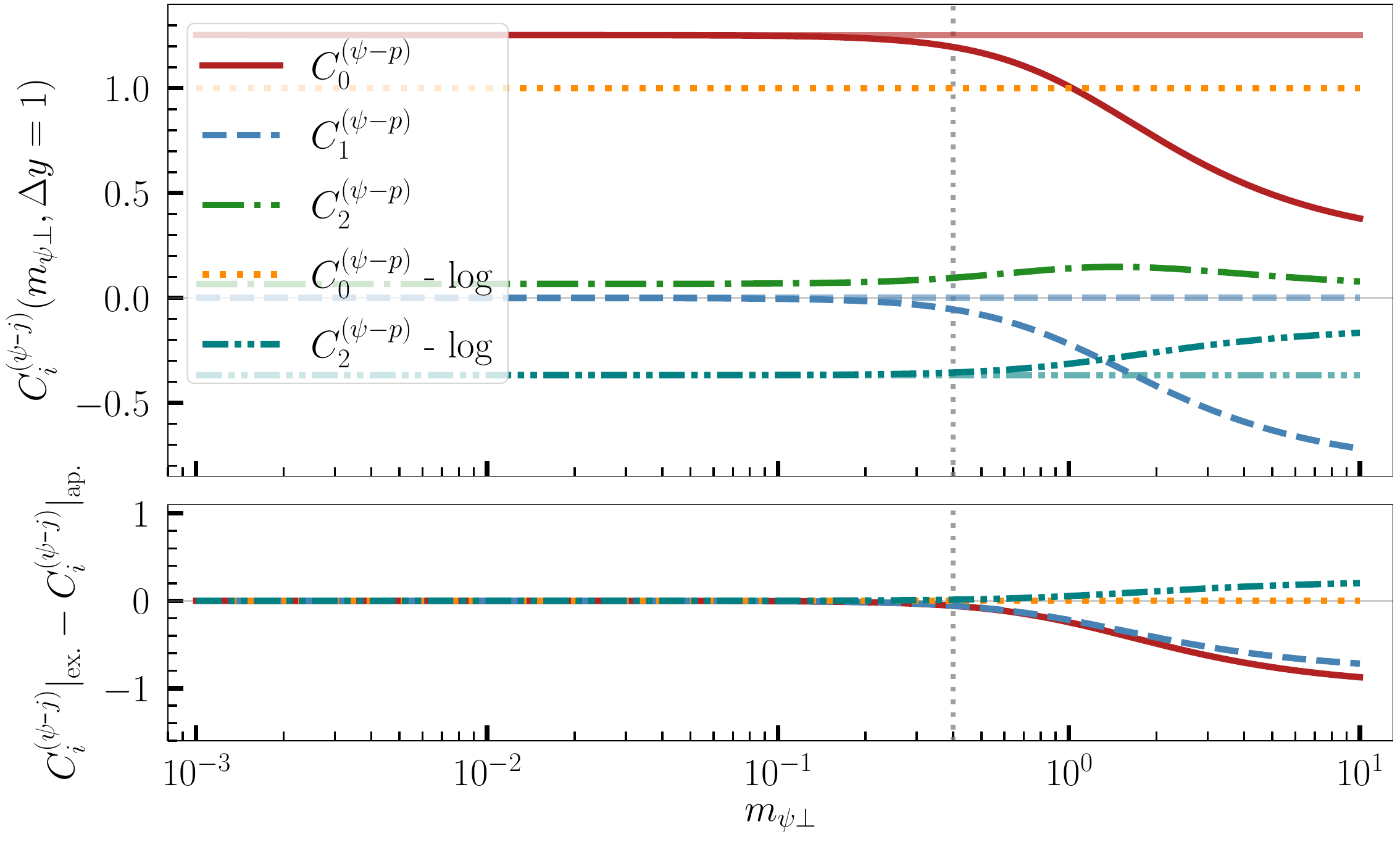}
\end{center}
\vskip -0.4cm \caption{\it  Same as Fig.~\ref{fig: coeff Jpsi 2 (Dy=0)} but for $\Delta y = 1$. At variance with the previous figure, we have that the coefficients include extra logarithms (see Eq.~\eqref{eq: I_psi2 approx Fourier series}), which are shown separately.}
\label{fig: coeff Jpsi 2 (Dy=1)}
\end{figure}

Among these three distributions, Eq.~\eqref{eq: I_psi2} is the most interesting. 
It does not depend on the jet variable $R$, since the integration of $\widehat S_g(k_j, k_\psi)$ is continuous at $\phi = 0$. However, it presents an additional dependence on the rapidity difference $\Delta y$, which affects its harmonic expansion coefficients, given by
\begin{align}
    I_{\psi \text{-} j}(m_{\psi\perp}, \Delta y, \phi)
    & = C_{0}^{(\psi \text{-} j)}(m_{\psi \perp}, \Delta y)  \nonumber\\
    & \phantom{=} + 2\, \sum_{n=1}^\infty C_{n}^{(\psi \text{-} p)} (m_{\psi \perp}, \Delta y)\, \cos(n \phi)\ .
\label{eq: I_psi2 Fourier series}
\end{align}
In particular, depending on the value of $\Delta y$ we can have additional logarithms of $\hat u/\hat t$ within the coefficients.
Moreover, the closer the two outgoing particles are to the production axis (namely $|\Delta y| \to \infty$), the less relevant the angular distribution in $I_{\psi \text{-} j} (m_{\psi\perp}, \Delta y, \phi)$ becomes, with its sole contribution being restricted to a logarithm of $\hat u/\hat t$. 
To see this effect, we consider the analytical expansion in the small-$m_{\psi\perp}$ limit for two values of $\Delta y$, namely $\Delta y = 0$ and $\Delta y = 1$:
\begin{subequations}
\begin{align}
    & I_{\psi \text{-} j}(m_{\psi\perp}, \Delta y = 0, \phi) \nonumber\\ &~~ = 2 \ln (4) - 4 \cos(2\phi) \left(\ln(4) - 1 \right) + \cdots
\end{align}
and
\begin{align}
    & I_{\psi \text{-} j}(m_{\psi\perp}, \Delta y = 1, \phi) \nonumber\\ &~~ = 2 \left[ 2 \Big(\ln(1 + {\rm e}) - 1 \Big) - \ln\frac{\hat u}{\hat t} - \frac{1}{2}\ln\frac{1 - M_\psi^2/\hat u}{1 - M_\psi^2/\hat t} \right]  \nonumber\\ & \phantom{=}~~
        - \frac{4}{\rm e} \cos(2\phi) \Bigg[ (1 + {\rm e}^2) \Big( \ln(1 + {\rm e}) - 1 \Big)  \nonumber\\ & \phantom{=}~~\quad + {\rm e} - \ln\frac{\hat u}{\hat t} - \frac{1}{2}\ln\frac{1 - M_\psi^2/\hat u}{1 - M_\psi^2/\hat t}\Bigg] + \dots \ .  
\end{align}
\label{eq: I_psi2 approx Fourier series}
\end{subequations}
Note that at small-$m_{\psi\perp}$ the distribution only contributes to the even modes of the Fourier expansion.
The complete dependence of the first coefficients of Eq.~\eqref{eq: I_psi2 Fourier series} for the same values of $\Delta y$ is shown in Figs.~\ref{fig: coeff Jpsi 2 (Dy=0)} and~\ref{fig: coeff Jpsi 2 (Dy=1)}, where once again it is shown that the approximation in Eq.~\eqref{eq: I_psi2 approx Fourier series} holds up to $m_{\psi\perp} \sim 0.4$}.

\textit{\textbf{Integral of $\bm{S_g(k_\psi, k_\psi)}$.}}
The function $S_g(k_\psi,k_\psi)$, nonzero only for massive particles, is given by
\begin{align}
    & S_g(k_\psi,k_\psi) \nonumber\\
    & \quad \approx \frac{2}{|\vec k_{g\perp}^2|^2} \frac{m_{\psi\perp}^2}{\Big(\sqrt{1 + m_{\psi\perp}^2}\cosh(\Delta y_{g\psi}) + \cos(\phi)\Big)^2} \ .
\end{align}
The azimuthal distribution arising from $S_g(k_\psi, k_\psi)$ is identify by 
\begin{align}
    & I_{\psi \text{-}\psi} (m_{\psi\perp}, \phi) \nonumber\\
    & \quad = 
    \int {\rm d} \Delta y_{g\psi}\, \frac{m_{\psi\perp}^2}{\Big(\sqrt{1 + m_{\psi\perp}^2}\cosh(\Delta y_{g\psi}) + \cos(\phi)\Big)^2}\ ,
\label{eq: I_psi3}
\end{align}
Moreover, when evaluated within the jet region, $S_g(k_\psi, k_\psi)$ generates
\begin{align}
    & I_{\psi \text{-}\psi}^{\rm jet} (R, m_{\psi\perp}, \Delta y, \phi) \nonumber\\
    & \quad = \frac{| \vec k_{g\perp}|^2}{2} \int {\rm d} \Delta y_{g}\, S_g(k_\psi, k_\psi)\, \Theta (\Delta_{k_g k_j} < R^2) \ .   
\end{align}

\begin{figure}[t]
\begin{center}
\includegraphics[width=.9\linewidth]{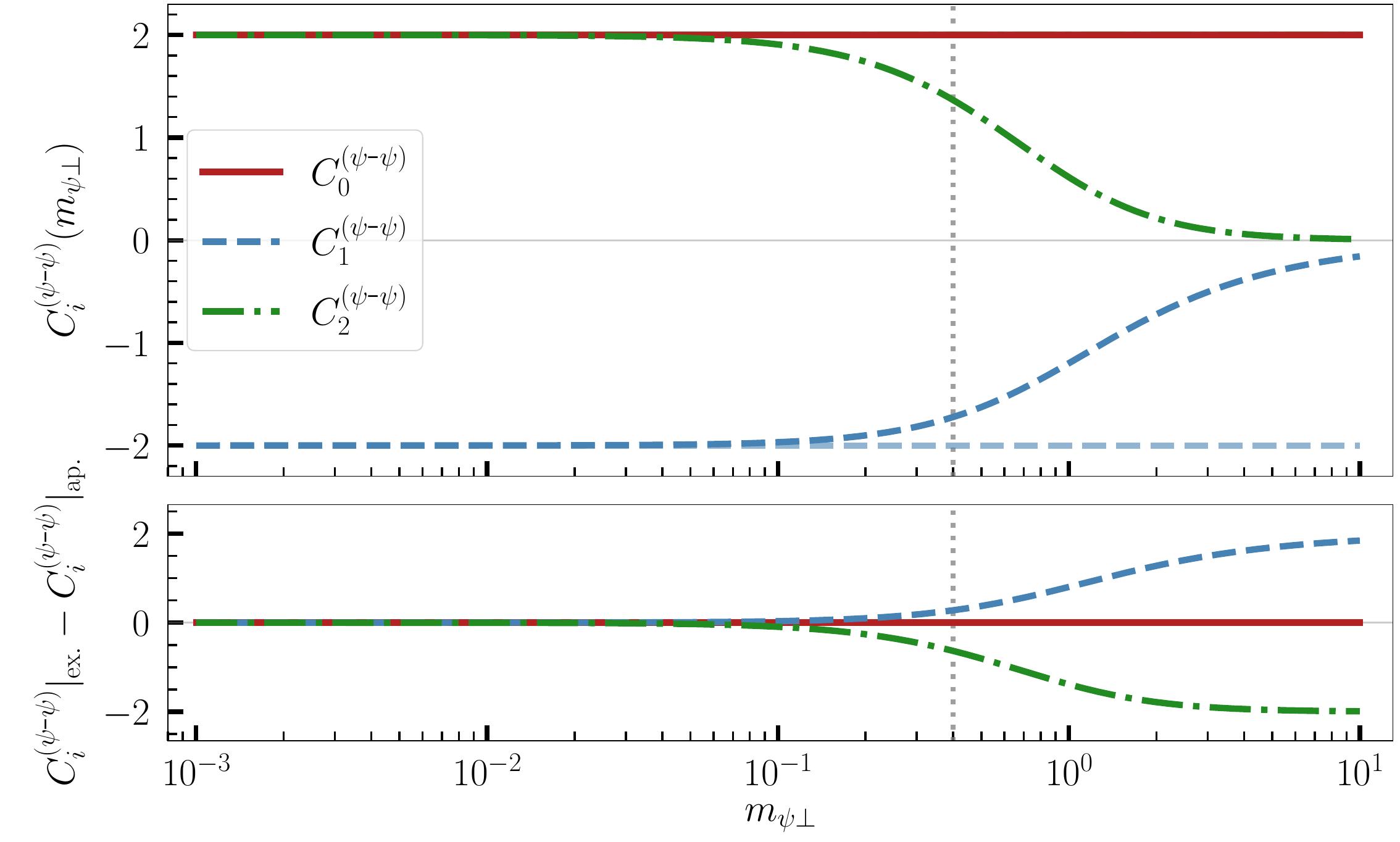}
\end{center}
\vskip -0.4cm \caption{\it Dependence of the first coefficients of Eq.~\eqref{eq: I_psi3 Fourier series} with respect to $m_{\psi\perp}$. Note that the coefficient $C_0$ does not vary with $m_{\psi\perp}$. Vertical dotted line corresponds to $m_{\psi\perp} = 0.4$. Panels follow the same logic as Fig.~\ref{fig: coeff jet}.}
\label{fig: coeff Jpsi 3}
\end{figure}

Also in this case, we perform the harmonic expansion
\begin{align}
    & I_{\psi \text{-}\psi} (m_{\psi\perp}, \phi) \nonumber\\
    & \quad = C_{0}^{(\psi \text{-} \psi)}(m_{\psi\perp}) + 2 \sum_{n=1}^\infty C_{n}^{(\psi \text{-} \psi)}(m_{\psi\perp})\, \cos(n \phi)\ .
\label{eq: I_psi3 Fourier series}
\end{align}
It is interesting to notice that these coefficients are nonzero for all $m_{\psi\perp}$. In particular, we have residual contributions in the small-$m_{\psi\perp}$ limit, with ${C_{0}^{(\psi \text{-} \psi)},\ C_{2}^{(\psi \text{-} \psi)} \to 2}$ and ${C_{1}^{(\psi \text{-} \psi)} \to - 2}$, which is expected due to the singular behavior of $I_{\psi \text{-} \psi}$ at $\phi = \pi$. 

For completeness, Fig.~\ref{fig: coeff Jpsi 3} shows the exact dependence of the coefficients on $m_{\psi\perp}$. From this figure we understand that $C_{0}^{(\psi \text{-} \psi)}$ is independent of $m_{\perp\psi}$, whereas $C_{1}^{(\psi \text{-} \psi)}$ and $C_{2}^{(\psi \text{-} \psi)}$ are negligible when $m_{\perp\psi} \gtrsim 1$.

\textit{\textbf{Overall distribution.}}
Combining the above derivations, we obtain the soft gluon radiation contributions in the CO channel
\begin{align}
    & \int \frac{{\rm d}^3k_g}{(2\pi)^3 2 E_{k_g}}\, 
    |\overline{{\cal A}_1^{g, (8)}}|^2\, \delta^{(2)} (q_\perp+k_{g\perp})\nonumber \\ 
    &~~ = \frac{\alpha_s C_A}{2\pi^2 |\vec q_\perp|^2} |\overline{{\cal A}_0^{g, (8)}}|^2\,  \Bigg[\ln\frac{\hat s}{|\vec q_\perp|^2} + \frac{1}{2}\ln \frac{1 - M_\psi^2/\hat u}{1 - M_\psi^2/\hat t}\nonumber\\
    &~~\phantom{=}\quad + I^g(R, m_{\psi\perp}, \Delta y, \phi)) \Bigg] + \textit{mixing}  \ ,
\label{eq: int CO}
\end{align}
for the gluon channel, and
\begin{align}
    & \int \frac{{\rm d}^3k_g}{(2\pi)^3 2 E_{k_g}}\, 
    |\overline{{\cal A}_1^{q, (8)}}|^2\, \delta^{(2)} (q_\perp+k_{g\perp})\nonumber \\ 
    &~~ = \frac{\alpha_s C_F}{2\pi^2 |\vec q_\perp|^2} |\overline{{\cal A}_0^{q, (8)}}|^2\, \Bigg[ \ln\frac{\hat s}{|\vec q_\perp|^2} + \frac{C_F - C_A}{C_F} \ln \frac{\hat t}{\hat u} 
    \nonumber\\
    &~~\phantom{=} + 
    \frac{C_A}{C_F}\,\bigg(\frac{1}{2} \ln \frac{1 - M_\psi^2/\hat u}{1 - M_\psi^2/\hat t} + \frac{C_A}{C_F}\, I^q(R, m_{\psi\perp}, \Delta y, \phi) \bigg) \Bigg]
    \nonumber\\
    &~~\phantom{=} + \textit{mixing}  \ ,
\label{eq: int CO - quark}
\end{align}
for the quark one.
The first term of Eqs.~\eqref{eq: int CO} and~\eqref{eq: int CO - quark} corresponds to the leading, double logarithmic behavior at low $q_\perp$, which is the same as the CS case. This implies that the soft gluon emission from the (massive) quarkonium does not provide double logarithms, a conclusion in line with other works~\cite{Sun:2012vc, Zhu:2012ts, Zhu:2013yxa, Echevarria:2019ynx, Fleming:2019pzj, Boer:2023zit}.
The other logarithms depend on the rapidities $y_j$ and $y_\psi$, where the sum $y_j + y_\psi$ differs from zero due to the presence of the $J/\psi$ mass.
The third term, $I^g$ for gluons and $I^q$ for quark, is the overall azimuthal distribution. More specifically, we have 
\begin{align}
    I^g(R, m_{\psi\perp}, \Delta y, \phi) & = I_j(R, \phi) + I_{\psi}(m_{\psi \perp}, \phi) \nonumber\\
    & \phantom{=} + \frac{1}{2} I_{\psi \text{-} j}(m_{\psi \perp}, \Delta y, 2\phi) \nonumber\\
    & \phantom{=} - \frac{1}{2} I_{\psi}^{\rm jet}(R, m_{\psi \perp}, \Delta y, \phi) \ ,
\label{eq: azimuth func gluon}
\end{align}
and
\begin{align}
    I^q(R, m_{\psi\perp}, \Delta y, \phi) & = \frac{C_F}{C_A} I_j(R, \phi) + I_{\psi}(m_{\psi \perp}, \phi) \nonumber\\
    & \phantom{=} + \frac{1}{2} I_{\psi \text{-} j}(m_{\psi \perp}, \Delta y, 2\phi) \nonumber\\
    & \phantom{=} - \frac{1}{2} I_{\psi}^{\rm jet}(R, m_{\psi \perp}, \Delta y, \phi) \ .
\label{eq: azimuth func quark}
\end{align}
Note that they differ only for the prefactor of $I_j$, namely the azimuthal dependence associated with the jet found in Eq.~\eqref{eq: int CS}.
The last three azimuthal distributions, namely ${I_\psi = I_{\psi \text{-} p} (m_{\psi\perp}, \phi) - \frac{1}{2} I_{\psi \text{-} \psi} (m_{\psi\perp}, \phi)}$, $I_{\psi \text{-} j}$ and $I_\psi^{\rm jet}$, are the novel terms due to the production of a CO state, with each $I_{\psi \text{-} v}$ generated from the corresponding $S_g(k_\psi, v)$.
Again, we expand them in terms of $\cos(n\phi)$ harmonics according to
Following the expansion of each term discussed above, we expand the overall distribution according to
\begin{equation}
    I^a (K, \phi) = 2 \sum_{n=0}^\infty C_n^{a} (K)\, \cos(n \phi)\ ,
\label{eq: Fourier Series CO}
\end{equation}
where $K$ is a shorthand notation for the dependence on $R$, $m_{\psi\perp}$ and $\Delta y$.

\subsection{Soft gluon emission and LDME evolution}
\label{sec: ldme mix}

\begin{figure}[t!]
\begin{center}
{\includegraphics[width=.5\linewidth, keepaspectratio]{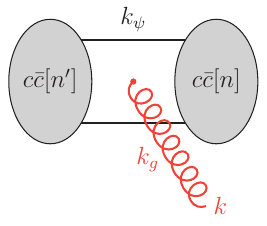}}
\end{center}
\vskip -0.4cm \caption{Soft gluon emission from the $c \bar c$ pair. We consider $n = {}^3 P_J^{(d)}$, while $n^\prime \equiv {}^3 P_J^{(d^\prime)}$, $n^\prime = {}^3 S_1^{(1)}$ or $n^\prime = {}^3 S_1^{(d^\prime)}$. The soft gluon, which carries out the color-index $k$, is given in red and it connects to both $c$ and $\bar c$.}
\label{fig: soft gluon emission cc}
\end{figure}

In this section, we present more details regarding the LDME mixing.
This mixing is generated from the evolution of a lower Fock-state to another one with higher quantum numbers, e.g.,~$S \to P$.
Being a characteristic of quarkonia, this feature can be observed only in the interference with soft gluons emitted from the $c \bar c$ pair (Fig.~\ref{fig: soft gluon emission cc}.) In particular, if one includes states up to a relative power $v^4$ in the NRQCD expansion, the contributions of the evolution in $J/\psi$ yields can only be observed in the $^3 P_J^{(8)}$ wave. In this case, the soft gluon emission can be decomposed into two terms, namely ${A_1^{(^3 P_J^{(8)})} = A_{1f}^{(^3 P_J^{(8)})} + A_{1d}^{(^3 P_J^{(8)})}}$ with
\begin{equation}
\begin{aligned}
    A_{1f}^{(^3 P_J^{(8)})} & = (i\, g_s\, f_{d d' k})\, \frac{k_\psi \cdot \epsilon_{\lambda_g}}{k_\psi \cdot k_g}\, A_0^{(^3 P_J^{(8)})}\ ,\\
    A_{1d}^{(^3 P_J^{(8)})} & = \left( -4 \sqrt{3} i\, g_s\, \frac{R_1^\prime}{R_0} \right) \frac{\epsilon_{L_z} \cdot \epsilon_{\lambda_g}}{k_\psi \cdot k_g}\\ & \phantom{=}\times \left( \sqrt{\frac{2}{N}}\, \delta_{dk}\, A_0^{(^3 S_1^{(1)})} + d_{d d' k}\,A_0^{(^3 S_1^{(8)})}\right) ,
\end{aligned}
\end{equation}
where $d$ is the color of the $c \bar c$, $k$ is the color of the soft gluon, and the amplitudes $A_0^{n}$ include the proper radial wave functions.
These two contributions do not interfere with each other, due to the orthogonality of the $J/\psi$ polarization vector, ${\epsilon_{L_z} \cdot k_\psi = 0}$. Moreover, the color structure of the studied process cancels the interference between the $A_{1d}^{(^3 P_J^{(8)})}$ emission with either the incoming or outgoing partons.
Consequently, this mixing term contributes only to the self-interference of soft gluons emitted by the $c \bar c$ via
\begin{align}
    |\overline{A_{1d}^{a, (^3 P_J^{(8)})}}|^2 & = 96\, g_s^2\, \frac{|R_1^\prime|^2}{M_\psi^2 |R_0|^2}\, \Big( C_F\, |\overline{A_{0}^{a, (^3 S_1^{(1)})}}|^2 \nonumber\\
    & \phantom{=} + B_F\, |\overline{A_{0}^{a, (^3 S_1^{(8)})}}|^2 \Big)\, S_g(k_\psi, k_\psi)\ ,
\label{eq: ldme evolution - amplitude}
\end{align}
with $B_F = (N_c^2 - 4)/4N_c$. Eq.~\eqref{eq: ldme evolution - amplitude}, summed over all partons ${a = g, q, \bar q}$, corresponds to the “\textit{mixing}” term of Eq.~\eqref{eq: A1 CO}.
Finally, upon integration over the phase space of the emitted soft gluon, one gets
\begin{align}
    & \int \frac{{\rm d}^3k_g}{(2\pi)^3 2 E_{k_g}}\, 
    |\overline{{\cal A}_{1d}^{a, (^3 P_J^{(8)})}}|^2\, \delta^{(2)} (q_\perp + k_{g\perp})\nonumber \\ 
    &~~ = \frac{\alpha_s}{2 \pi^2 |\vec q_\perp|^2} \frac{96}{M_\psi^2} \left( \frac{ |\overline{A_{0}^{a, (^3 S_1^{(1)})}}|^2}{\langle {\cal O}_1 (^3 S_1) \rangle} + B_F\,  \frac{ |\overline{A_{0}^{a, (^3 S_1^{(8)})}}|^2}{\langle {\cal O}_8 (^3 S_1) \rangle}  \right) \nonumber \\ 
    &~~ \phantom{=} \times \langle {\cal O}_8 (^3 P_0) \rangle \frac{I_{\psi \text{-} \psi}(m_{\psi\perp}, \phi) - I^{\rm jet}_{\psi \text{-} \psi}(R, m_{\psi\perp}, \Delta y, \phi)}{2} ,
\label{eq: ldme evolution - integration}
\end{align}
where we have taken into account the relation between the radial functions and the LDME with the proper normalization factors. Eq.~\eqref{eq: ldme evolution - integration}, which is in agreement with \cite{Butenschoen:2019lef}, corresponds to the “\textit{mixing}” term reported in Eq.~\eqref{eq: int CO}, with $a = g$, and Eq.~\eqref{eq: int CO - quark}, with $a = q$.

Finally, note that to get the contributions of the LDME evolution to Eq.~\eqref{eq: diff cross-section soft gluon radiation}, we have performed the expansion in harmonics of the azimuthal distribution and explicitly taken the relation between the CO and CS gluonic channels 
${|\overline{A_{0}^{g, (^3 S_1^{(8)})}}|^2 = \frac{15}{8}\frac{\langle {\cal O}_8 (^3 S_1) \rangle}{\langle {\cal O}_1 (^3 S_1) \rangle}\, |\overline{A_{0}^{g, (^3 S_1^{(1)})}}|^2}$, while ${|\overline{A_{0}^{q, (^3 S_1^{(1)})}}|^2 = 0}$.

\subsection{Summary of one-loop results}

Summarizing the above results, the differential cross section including the soft gluon radiation in the correlation limit is given by
\begin{widetext}
\begin{align}
    \frac{d^4\sigma^{(c)}}{d\Omega} = 
    \frac{\alpha_s C_A}{2\pi^2 |\vec q_\perp|^2}\, x & \Bigg\{ \sigma_0^{g, (c)}\,f_g(x)\, \Bigg[\ln\frac{\hat s}{|\vec q_\perp|^2} 
    + 2 \sum_{n = 0}^\infty C_n^{g, (c)}(R,m_{\psi\perp}, \Delta y = 0)\, \cos(n \phi)\Bigg] 
    \nonumber\\
    & \phantom{=}~ 
    + \sum_q \sigma_0^{q, (c)}\,f_q(x)\, \Bigg[ \frac{C_F}{C_A}\, \ln\frac{\hat s}{|\vec q_\perp|^2} 
    + 2 \sum_{n = 0}^\infty C_n^{q, (c)}(R,m_{\psi\perp}, \Delta y = 0)\, \cos(n \phi)\Bigg]
    \nonumber\\
    & \phantom{=}~ 
    + \delta_{c8}\, \frac{96}{M_\psi^2} \left( 1 + \frac{15}{8} B_F\right) \sigma_0^{g, (^3 S_1^{(1)})}\, \frac{\langle O_8 (^3 P_0) \rangle}{\langle O_1 (^3 S_1) \rangle}\,f_g(x) \sum_{n = 0}^\infty C_n^{\rm mix}(R, m_{\psi\perp}, \Delta y = 0)\, \cos(n \phi)
    \nonumber\\
    & \phantom{=}~ 
    + \delta_{c8}\, \sum_q \frac{96}{M_\psi^2}\, B_F\, \sigma_0^{q, (^3 S_1^{(8)})}\, \frac{\langle O_8 (^3 P_0) \rangle}{\langle O_8 (^3 S_1) \rangle}\,f_q(x) \sum_{n = 0}^\infty C_n^{\rm mix}(R, m_{\psi\perp}, \Delta y = 0)\, \cos(n \phi) \Bigg\}\ ,
\label{eq: diff cross-section soft gluon radiation}  
\end{align}
\end{widetext}
where $\sigma_0^{g, (c)}$ ($\sigma_0^{q, (c)}$), with $c = 1,\ 8$, represents the LO gluon (quark) cross section~\cite{Ko:1996xw}, ${{\rm d}\Omega \equiv {\rm d}y_j {\rm d}y_\psi {\rm d}^2 \vec P_\perp {\rm d}^2 \vec q_\perp}$ the phase space, and $f_g(x)$ is the gluon distribution. 
Note that the quarks solely contribute to the CO channel, as well as the LDME mixing, where for the dominant LDMEs in $J/\psi$ production we need to include the evolution from the $^3 S_1^{(c)}$ states, both CS and CO.
We are considering the case where $y_j = y_\psi = 0$ for simplicity since this scenario does not present additional, nondivergent logarithms. Hence, in Eq.~\eqref{eq: diff cross-section soft gluon radiation} we have that the first and third terms correspond to the leading double logarithms, while the second and fourth lines stand for the single logarithms. Since the latter carry the azimuthal dependence, we have already expanded them into harmonics, with the coefficients $C_n^{a, (c)}$ connected to those in the Fourier series of each distribution.
Therefore they depend on the jet size $R$, the heavy quarkonium mass via ${m_{\psi\perp} = M_\psi/P_\perp}$ and the rapidity difference ${\Delta y = y_\psi - y_j}$. 
Most importantly, these coefficients depend crucially on the production channel considered.

\begin{figure*}[t!]
\begin{center}
\subfloat[\label{fig: heatmap c0 g}]{\includegraphics[width=.35\linewidth, keepaspectratio]{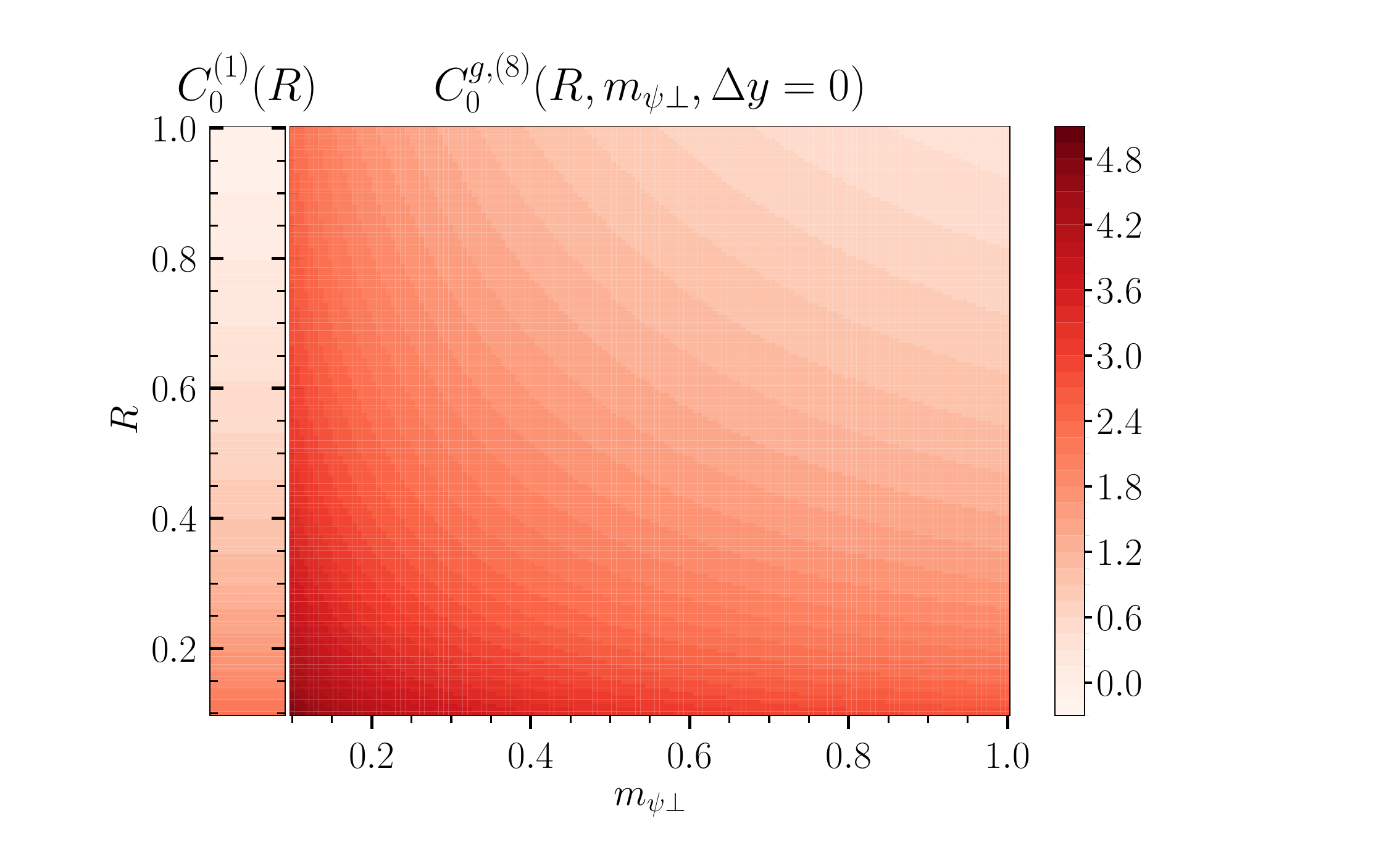}}
\subfloat[\label{fig: heatmap c1 g}]{\includegraphics[width=.35\linewidth, keepaspectratio]{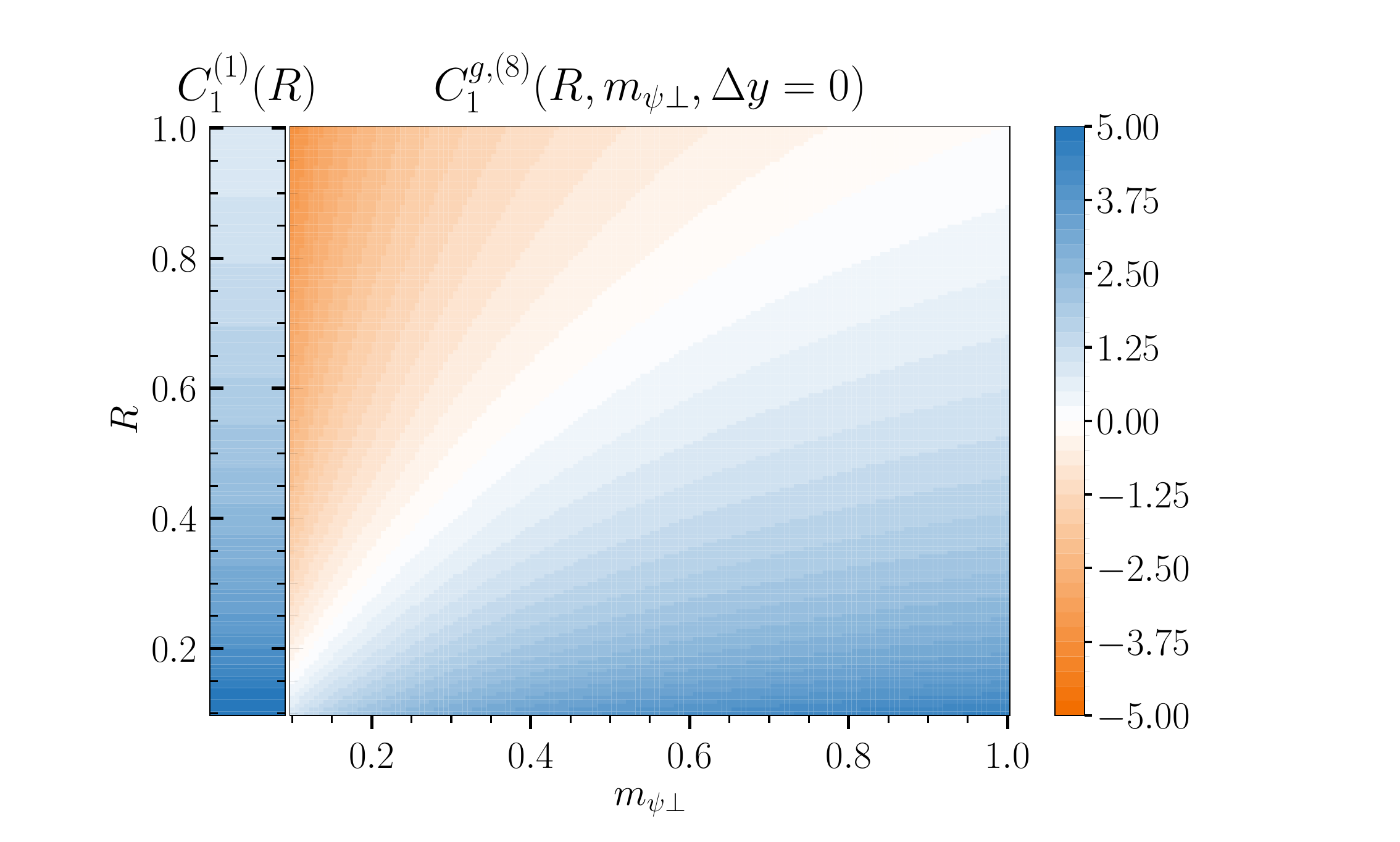}}
\subfloat[\label{fig: heatmap c2 g}]{\includegraphics[width=.35\linewidth, keepaspectratio]{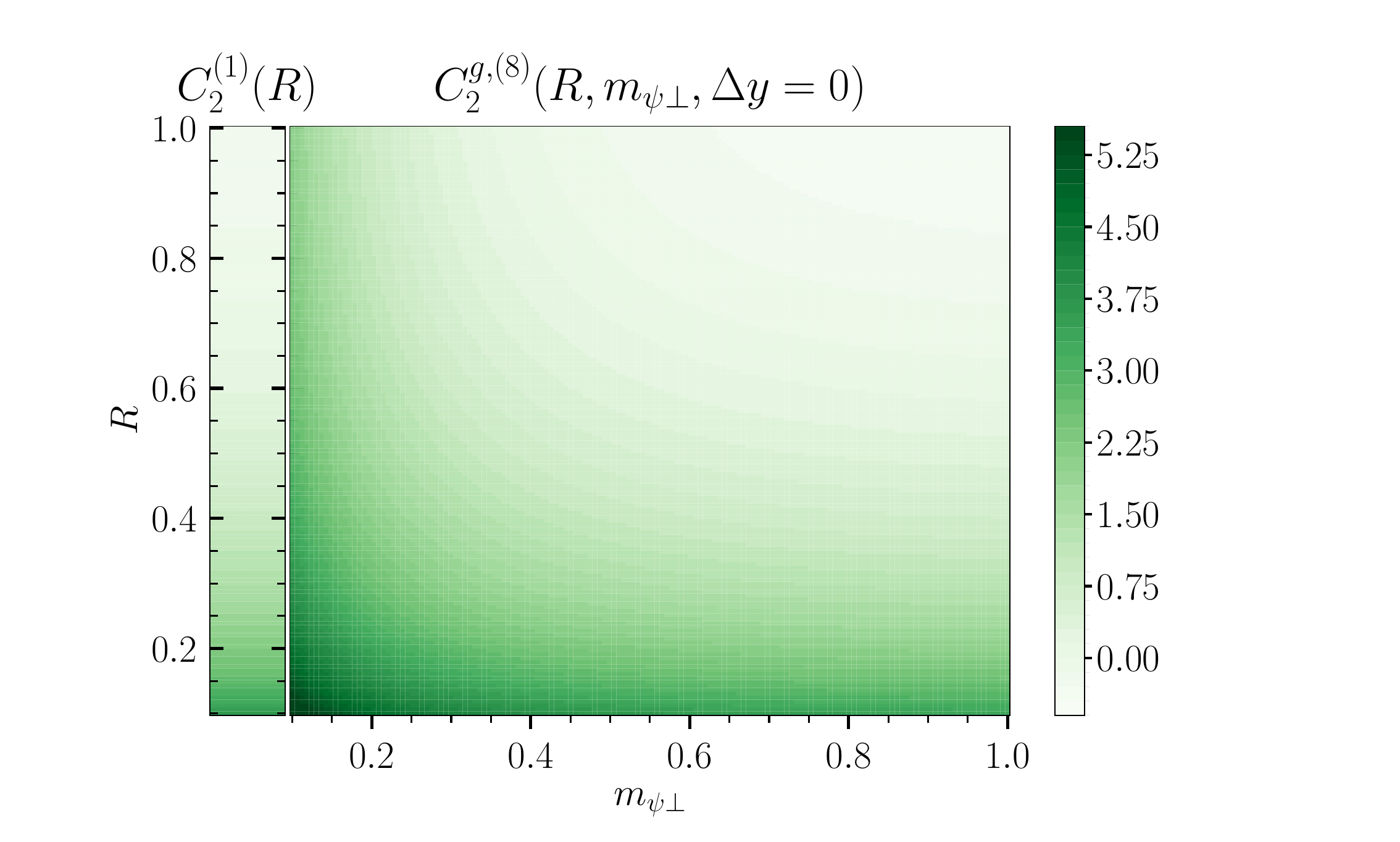}}
\end{center}
\vskip -0.4cm \caption{\it Dependence of the first coefficients on the jet size $R$ and the variable $m_{\psi\perp}$ for CO production in the gluon channel and for transverse production ($\Delta y = 0$): (a) $C_0$, (b) $C_1$, and (c) $C_2$. A comparison with the CS channel, for which coefficients solely depend on $R$, is given by the column on the left.}
\label{fig: heatmap CO coefficients (gluon)}
\end{figure*}

\begin{figure*}[t!]
\begin{center}
\subfloat[\label{fig: heatmap c0 q}]{\includegraphics[width=.35\linewidth, keepaspectratio]{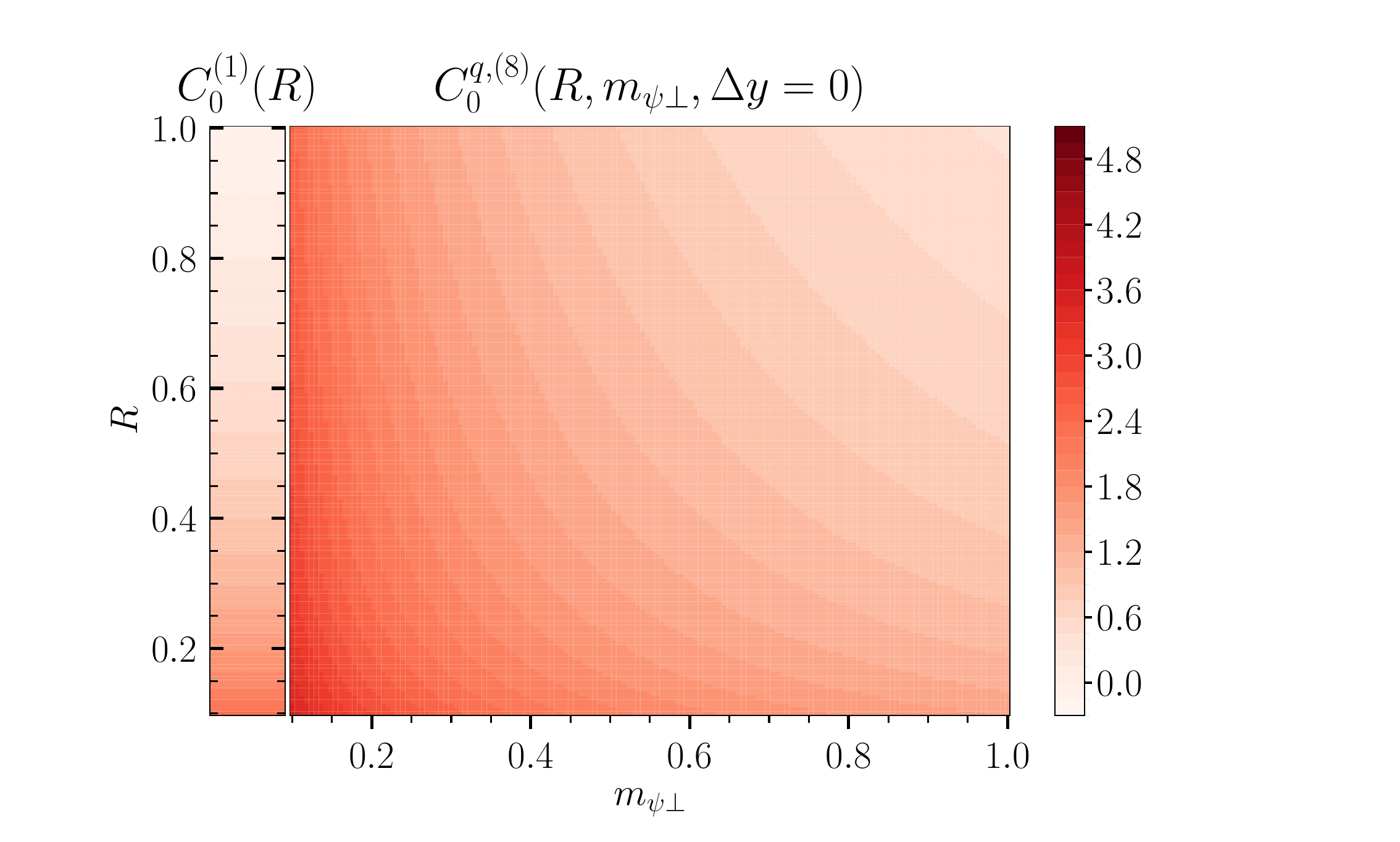}}
\subfloat[\label{fig: heatmap c1 q}]{\includegraphics[width=.35\linewidth, keepaspectratio]{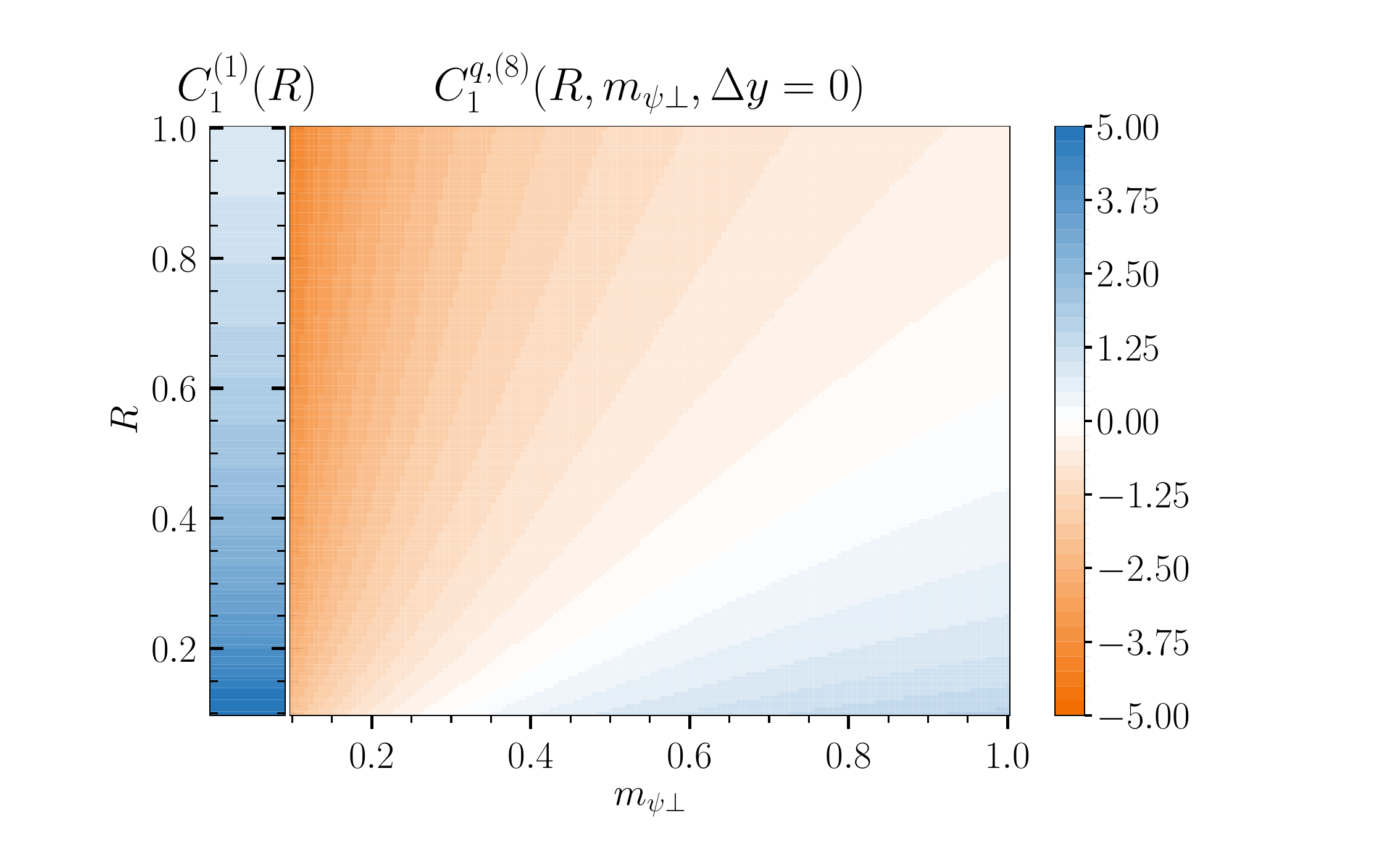}}
\subfloat[\label{fig: heatmap c2 q}]{\includegraphics[width=.35\linewidth, keepaspectratio]{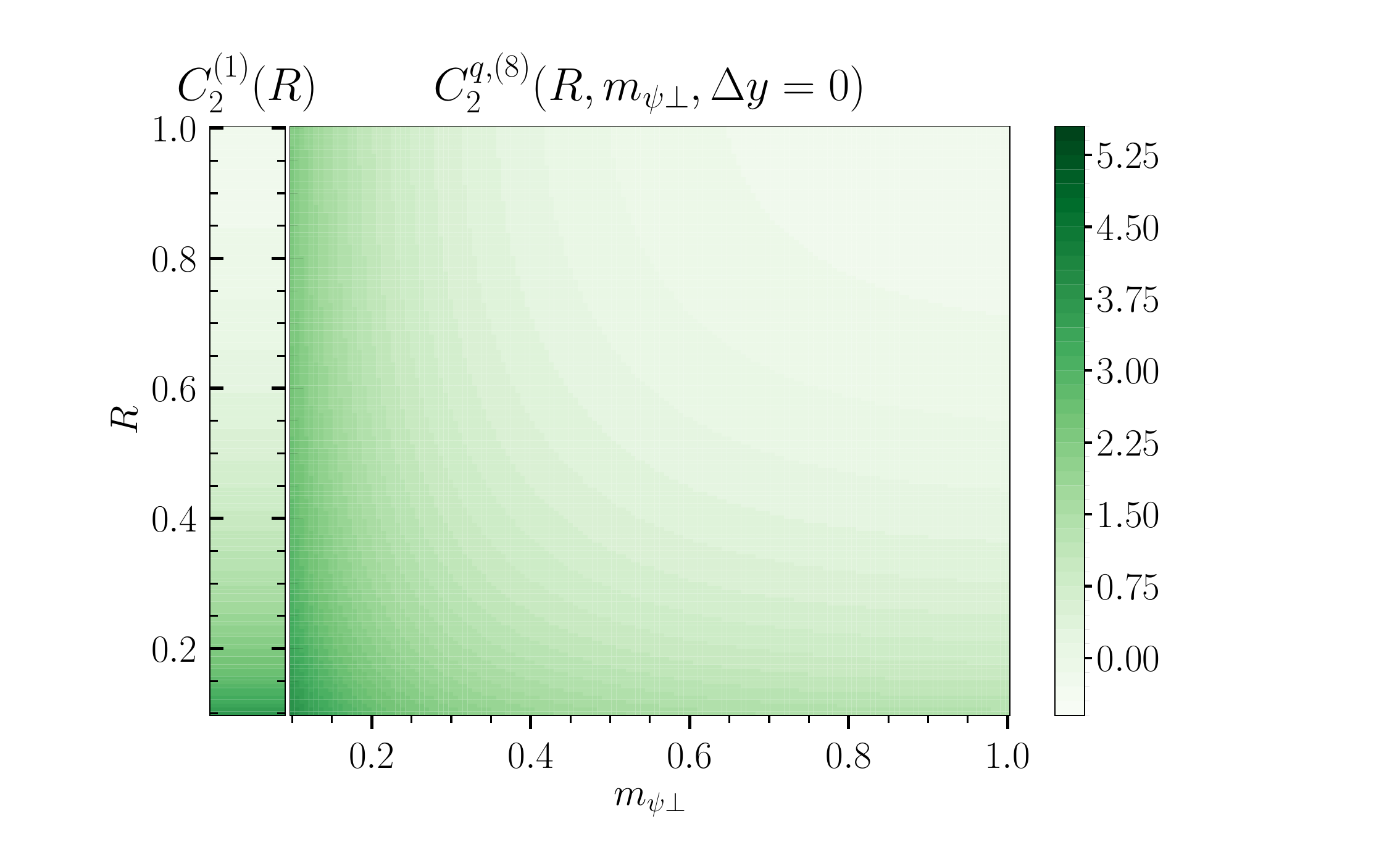}}
\end{center}
\vskip -0.4cm \caption{\it Same as Fig.~\ref{fig: heatmap CO coefficients (gluon)} but for the coefficients occurring in the quark induced process.}
\label{fig: heatmap CO coefficients (quark)}
\end{figure*}

To illustrate these differences, we consider a typical kinematic for future EIC measurements.
We take ${R = 0.4}$ and ${m_{\psi\perp} = 0.26}$, that corresponds to ${k_{j\perp} \approx 12}$~GeV. Within this choice, the first three coefficients of the Fourier expansion in Eq.~\eqref{eq: diff cross-section soft gluon radiation} are given in Table~\ref{tab: coefficients}, displaying a clear difference between CS and CO channels.
In particular, the coefficient $C_1^{(1)}$ is significantly positive due to the soft gluon radiation associated with the jet in the final state. However, the soft gluon radiation associated with heavy quark pair has an opposite sign.
Thus, a proper choice of $m_{\psi\perp}$ can reduce the magnitude of $C_1^{(8)}$ 
when compared to $C_1^{(1)}$, while on the other hand $C_2^{(8)}$ and $C_2^{(1)}$ stay of more or less the same order. 
Beyond the $m_{\psi\perp}$ used in Table~\ref{tab: coefficients}, other values can also be considered, although they might be not accessible at the EIC.

\begin{table}[]
\centering
\resizebox{.65\columnwidth}{!}{
\begin{tabular}{|c|ccc|}
\toprule
\rule[-1.5ex]{0pt}{4.5ex} 
\textbf{Mechanism} & $\bm C_{\bm 0}^{\bm (\bm c \bm)}$ & $\bm C_{\bm 1}^{\bm (\bm c \bm)}$ & $\bm C_{\bm 2}^{\bm (\bm c \bm)}$ \\ \midrule
\rule[-1.2ex]{0pt}{4.ex} CS & $0.89$ & $2.61$ & $0.95$ \\ 
\rule[-1.2ex]{0pt}{4.ex} CO gluon 
& $2.45$ & $-0.08$ & $1.70$ \\
\rule[-1.2ex]{0pt}{4.ex} CO quark 
& $1.96$ & $-1.52$ & $1.17$ \\
\rule[-1.2ex]{0pt}{4.ex} CO mix
& $0.50$ & $-0.93$ & $0.81$ \\
\bottomrule
\end{tabular}
}
\caption{\it First coefficients of the azimuthal correlation Fourier expansions for the CS and CO mechanisms with $R = 0.4$ and $m_{\psi\perp} = 0.26$. Note that the coefficients within the CS channel are independent of $k_{j\perp}$.}
\label{tab: coefficients}
\end{table}

In Figs.~\ref{fig: heatmap CO coefficients (gluon)} and~\ref{fig: heatmap CO coefficients (quark)} we present a more comprehensive dependency of the first coefficients of the Fourier expansion in Eq.~\eqref{eq: diff cross-section soft gluon radiation} on both $R$ and $m_{\psi\perp}$ in the CO channel. Within the CS channel, the coefficients only depend on $R$ and are therefore given as columns.
Fig.~\ref{fig: heatmap CO coefficients (gluon)} corresponds to the gluon channel (first line of Eq.~\eqref{eq: diff cross-section soft gluon radiation}) while the Fig.~\ref{fig: heatmap CO coefficients (quark)} to the quark one (second line of Eq.~\eqref{eq: diff cross-section soft gluon radiation}).
Note that the LDME evolution is directly related to $I_{\psi \text{-} \psi}$, see Eq.~\eqref{eq: ldme evolution - integration}, so the comparison with the coefficients in the CS channel is already understood from Figs~\ref{fig: coeff jet} (CS) and~\ref{fig: coeff Jpsi 3} (CO LDME mix).
By comparing the color shades in the different rows, hence at fixed $R$, it is evident that predictions that include the CO mechanism are significantly different from those exclusively driven by the CS one.
This holds for both gluons and quarks, with the qualitative behaviors in the two partonic channels being, in fact, the same.
More specifically, we saw that in both cases the sign of the $C_1^{(8)}$ coefficient becomes opposite to that of $C_1^{(1)}$ for a sufficiently low $m_{\psi\perp}$ value. 
Of course, the specific $m_{\psi\perp}$ at which this occurs can vary depending on both $R$ and the partonic channel considered, with quarks consistently requiring higher $m_{\psi\perp}$ values (which corresponds to lower $|\vec k_{j\perp}|$ values) compared to gluons.
Thus, depending on the dominance of gluons over quarks or vice versa, we will have the same features as those discussed in Fig.~\ref{fig: asymmetries NRQCD} but seen at different values of $|\vec k_{j\perp}|$.
In addition, also the LDME mixing will not spoil this picture since, for $R<1$ and independently from $m_{\psi\perp}$, $C_1^{\rm mix}$ has an opposite sign to $C_1^{(1)}$, while $C_2^{\rm mix}$ and $C_2^{(1)}$ have the same sign.

\section{All order resummation and predictions for the EIC}
\label{sec: resummation and predictions}
All order resummation is needed to make reliable predictions for the soft gluon radiation contributions.
Following the standard TMD framework, we have
\begin{widetext}
\begin{align}
    \frac{d^4\sigma^{(c)}}{d\Omega} & = 
        \sigma_0^{g, (c)} \int\frac{{\rm d}\vec{b}_\perp^2}{4\pi}\, \bigg(J_0(|\vec b_\perp||\vec q_\perp|) + 2\, \frac{C_A \alpha_s}{n \pi}\,  J_n(|\vec b_\perp||\vec q_\perp|)\, C_n^{g, (c)}\, \cos(n \phi) \bigg)\,  \widetilde{W}_{0}^{g, (c)} (|\vec b_\perp|)
    \nonumber \\& \phantom{=} + 
        \sum_q \sigma_0^{q, (c)} \int\frac{{\rm d}\vec{b}_\perp^2}{4\pi}\, \bigg(J_0(|\vec b_\perp||\vec q_\perp|) + 2\, \frac{C_A \alpha_s}{n \pi}\,  J_n(|\vec b_\perp||\vec q_\perp|)\, C_n^{q, (c)}\, \cos(n \phi) \bigg)\,  \widetilde{W}_{0}^{q, (c)}(|\vec b_\perp|) 
    \nonumber \\ & \phantom{=} 
        + \delta_{c8}\, \sigma_0^{g, (^3 S_1^{(1)})}\, \frac{\langle O_8 (^3 P_0) \rangle}{\langle O_1 (^3 S_1) \rangle} \int\frac{{\rm d}\vec{b}_\perp^2}{4\pi}\, \bigg( J_0(|\vec b_\perp||\vec q_\perp|) 
    \nonumber \\ & \phantom{=} \hspace{5.5cm}
    + \frac{192}{M_\psi^2} \left( 1 + \frac{15}{8} B_F\right)\frac{C_A \alpha_s}{n \pi}\, J_n(|\vec b_\perp||\vec q_\perp|)\, C_n^{\rm mix}\, \cos(n \phi) \bigg) \, W_0^{g, {\rm mix}}(|\vec b_\perp|)
    \nonumber \\ & \phantom{=} 
        + \delta_{c8}\, \sum_q  \sigma_0^{q, (^3 S_1^{(8)})}\,  \frac{\langle O_8 (^3 P_0) \rangle}{\langle O_8 (^3 S_1) \rangle} \int\frac{{\rm d}\vec{b}_\perp^2}{4\pi}\, \bigg( J_0(|\vec b_\perp||\vec q_\perp|) 
    \nonumber \\ & \phantom{=} \hspace{6cm} + \frac{192}{M_\psi^2} \, B_F\, \frac{C_A \alpha_s}{n \pi}\, J_n(|\vec b_\perp||\vec q_\perp|)\, C_n^{\rm mix}\, \cos(n \phi) \bigg)\, W_0^{q, {\rm mix}}(|\vec b_\perp|)\ ,
\label{eq: resummed cross section}
\end{align}
\end{widetext}
where
\begin{equation}
    \widetilde{W}_0^{a, {\rm ch.}}(|\vec b_\perp|) = x\,f_a(x,\mu_b)
     \, {\rm e}^{-S^{a, {\rm ch.}}(P_\perp, b_\perp)}\ , 
 \label{eq: W0}
\end{equation}
with ${{\rm ch.} = (1),\ (2),\ {\rm mix}}$.
Note that in Eq.~\eqref{eq: resummed cross section} we have already included higher-order double logarithmic corrections present also for the angular dependent term~\cite{Catani:2014qha,Catani:2017tuc}.
Although we have derived Eq.~\eqref{eq: resummed cross section} at the one-loop order, it holds also when next-order contributions are included. In particular, real emissions will modify the coefficients of the double- and/or single-logarithms included in $W_0$, while virtual corrections apply to the hard factor $\sigma_0$. 
A consistent computation of all these next-order contributions in the correlation limit is still lacking and, for this reason, we will not include them in the following discussion. Note that, by excluding them we are missing corrections of ${\cal O} (q_\perp^2/k_{j\perp}^2)$, which can affect the high $q_\perp$ part of the spectrum and not the general result reported below.
The Sudakov form factor $S(P_\perp,b_\perp)$ is separated into perturbative and nonperturbative parts:
$S(P_\perp,b_\perp) = S_{\rm pert.}(P_\perp,b_\perp) + S_{\rm NP}(P_\perp,b_\perp)$. 
The perturbative parts at one loop are defined as
\begin{equation}
\begin{aligned}
S_{\rm pert.}^{g, (c)} & = \int^{\hat s}_{\mu_{b_*}^2} \frac{d\mu^2}{\mu^2} \frac{\alpha_s C_A}{2\pi}
\left[\ln\frac{\hat s}{\mu^2} - 2\, \beta_0 + 2\, C_0^{g, (c)}\right]\ , \\
S_{\rm pert.}^{q, (c)} & = \int^{\hat s}_{\mu_{b_*}^2} \frac{d\mu^2}{\mu^2} \frac{\alpha_s C_F}{2\pi}
\left[\ln\frac{\hat s}{\mu^2} - \frac{3}{2} + 2\,\frac{C_A}{C_F}\, C_0^{q, (c)}\right]\ ,
\end{aligned}
\label{eq: Sudakov pert.}
\end{equation}
while for the LDME evolution terms we have
\begin{equation}
\begin{aligned}
S_{\rm pert.}^{g, {\rm mix}} & = \int^{\hat s}_{\mu_{b_*}^2} \frac{d\mu^2}{\mu^2} \frac{\alpha_s C_A}{2\pi}
\Bigg[\ln\frac{\hat s}{\mu^2} - 2\, \beta_0  \\
& \phantom{=} \hspace{2.8cm} + \frac{192}{M_\psi^2} \left( 1 + \frac{15}{8} B_F\right)\, C_0^{\rm mix}\Bigg]\ , \\
S_{\rm pert.}^{q, {\rm mix}} & = \int^{\hat s}_{\mu_{b_*}^2} \frac{d\mu^2}{\mu^2} \frac{\alpha_s C_F}{2\pi}
\left[\ln\frac{\hat s}{\mu^2} - \frac{3}{2} + \frac{192}{M_\psi^2}\,\frac{C_A}{C_F}\, B_F\, C_0^{\rm mix}\right]\ ,
\end{aligned}
\label{eq: Sudakov pert. - LDME mix}
\end{equation}
where $\beta_0 = 11/12 - N_f/18$ and $\mu_{b} = b_0/b_\perp$ with $b_0 = 2\, {\rm e}^{\Gamma_E}$ and $\Gamma_E$ being the Euler's constant. 
Moreover, in Eq.~\eqref{eq: Sudakov pert.} we have already introduced the $b_*$-prescription~\cite{Collins:1984kg}, where $b_* = b_\perp/\sqrt{1 + (b_\perp/b_{\rm max})^2}$ with $b_{\rm max} = 1.5$~GeV.
For the nonperturbative part, we have contributions driven by the incoming parton and the outgoing jet and quarkonium. For the first one, we employ the nonperturbative Sudakov found for TMD quark distributions in Refs.~\cite{Su:2014wpa,Prokudin:2015ysa} with the appropriate Casimir scaling for gluons
\begin{equation}
\begin{aligned}
    S_{\rm NP}^{q} & = \left[ 0.106\, b_\perp^2 + 0.42 \ln\frac{P_{\perp}}{Q_0} \ln\frac{b_\perp}{b_*} \right] \ , \\
    S_{\rm NP}^{g} & = \frac{C_A}{C_F} \, S_{\rm NP}^{q}\ , 
\end{aligned}
\label{eq: Sudakov non-pert. (gluon)}
\end{equation}
where $Q_0^2 = 2.4$~GeV$^2$. 
For the others, we assume that the nonperturbative contribution associated with the jet is given by 
${S_{\rm NP}^{\rm jet} = g_\Lambda^{\rm jet}\, b_\perp^2}$,  
and the $J/\psi$ one by
${S_{\rm NP}^\psi = g_\Lambda^\psi\, b_\perp^2}$.
Overall, we employ  ${S_{\rm NP}^{(1)} = S_{\rm NP}^{g} + S_{\rm NP}^{\rm jet}}$  for the CS mechanism and ${S_{\rm NP}^{a, (8)} = S_{\rm NP}^{a}} + S_{\rm NP}^{\rm jet} + S_{\rm NP}^{\psi}$  for the CO channel. 
We take $g_\Lambda^{\rm jet} = 0.225~{\rm GeV}^2$, in line with \cite{Hatta:2020bgy, Hatta:2021jcd}, and $g_\Lambda^\psi = g_\Lambda^{\rm jet}$, upon the assumption that the CO final state generates a nonperturbative form factor similar to that of the (gluon) jet. Besides, we checked that the final azimuthal asymmetries do not depend significantly on these parameters, as expected.
From the above expressions, we find that the azimuthal asymmetries of $\cos(\phi)$ and $\cos(2\phi)$ are linearly proportional to the respective $C_1$ and $C_2$ coefficients which are different between the CS and CO channels. Therefore, they can be used to probe these two production mechanisms. 

\begin{figure}[t]
  \centering
  \includegraphics[width=\linewidth]{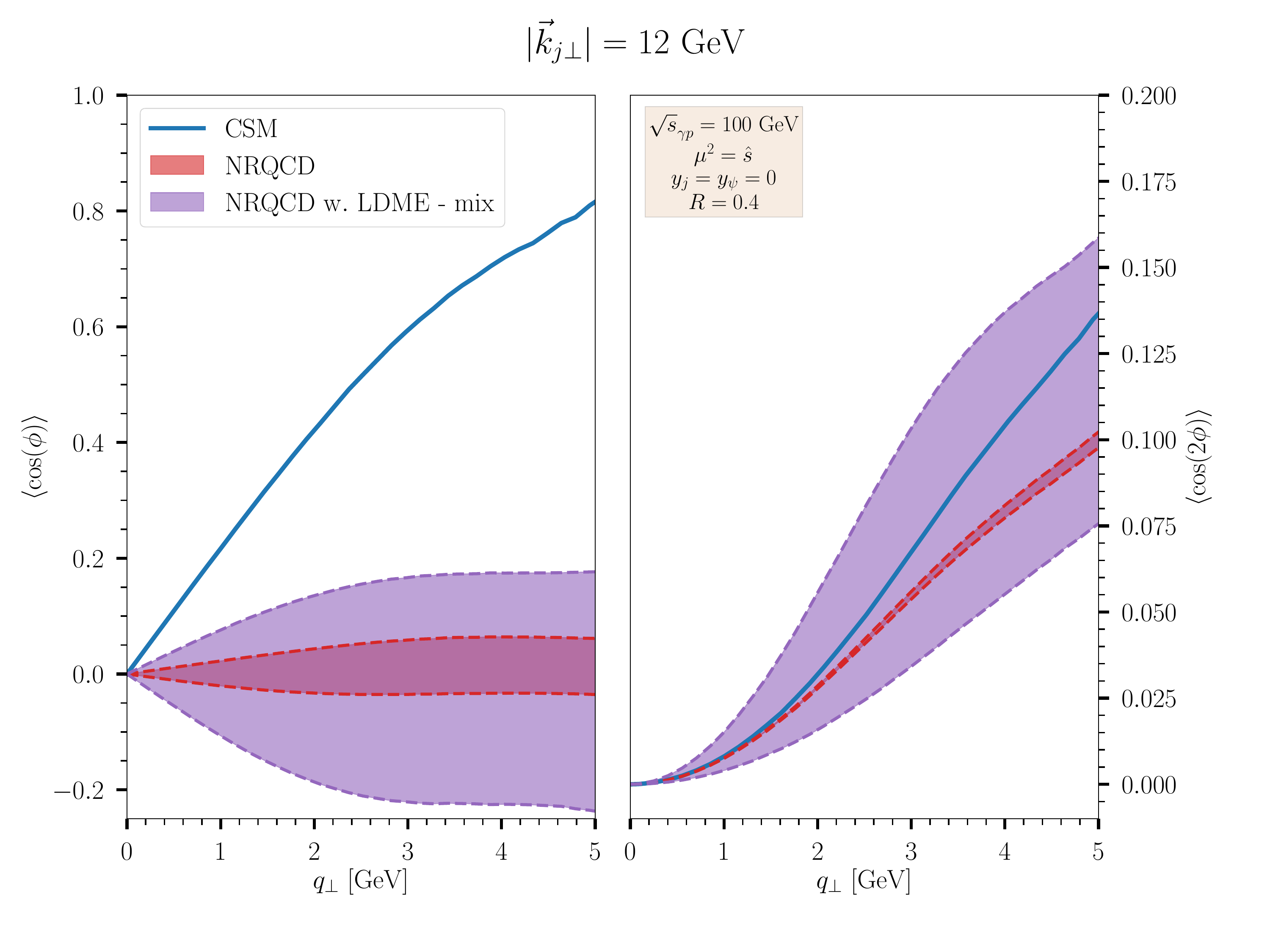}
\caption{\it Averaged resummed azimuthal asymmetries for $J/\psi$ plus jet photoproduction at $\sqrt s_{\gamma p} = 100$~GeV, $|\vec k_{j\perp}| = 12$~GeV and jet size $R = 0.4$. The solid blue line is the CSM predictions, whereas for the NRQCD approach we have the red and purple bands, where the latter includes the LDME evolution contribution. The bands are obtained by combining the results from different LDMEs central values.}
\label{fig: asymmetries NRQCD}
\end{figure}

As an example, in Fig.~\ref{fig: asymmetries NRQCD}, we show the numerical results for $\langle \cos(\phi)\rangle$ and $\langle \cos(2\phi)\rangle$ as functions of $q_\perp$ for realistic kinematics accessible at the EIC, with ${\sqrt s_{\gamma p} = 100~{\rm GeV}}$, $R = 0.4$ and ${|\vec k_{j\perp}| = 12~{\rm GeV}}$. For the parton distribution function, we have employed the MSHT20 LO set~\cite{Bailey:2020ooq}.
Predictions are shown for both the color-singlet model (CSM) and NRQCD approach, with the latter strongly depending on the relative fraction of the CS and total CO contributions. In particular we have included the following CO channels: $^1 S_0^{(8)}$, $^3 S_1^{(8)}$ and $^3 P_J^{(8)}$ Fock states in both gluon and quark channels.
The bands of the NRQCD predictions are constructed by combining several central values of the associated LDMEs from different global analyses~\cite{Butenschoen:2011yh,Chao:2012iv,Gong:2012ug,Sharma:2012dy,Brambilla:2022rjd}; among them, we remark that solely \cite{Butenschoen:2011yh} agrees with HERA photoproduction data. 
The red band is obtained by neglecting the LDME evolution contributions, with the CO azimuthal distribution thence independent of the CO Fock state. In this scenario, the result is mainly driven by the gluon channel.
Upon the inclusion of the LDME evolution, which is found non-negligible only for the quark channel for the kinematics explored in Fig.~\ref{fig: asymmetries NRQCD}, the band size increases significantly. This large uncertainty offers, with precise enough data, the additional opportunity to utilize these observables in LDME fits to greatly constrain their values.
Both NRQCD bands are mostly dominated by the CO mechanism, and we expect that the inclusion of higher-order states (in $v$) does not significantly affect the result shown in Fig.~\ref{fig: asymmetries NRQCD}.
We also expect that the following conclusions, based on Fig.~\ref{fig: asymmetries NRQCD}, apply to other kinematic regions. Thus, we believe that our findings are broader than the specific scenario discussed here.

The resummed form of $\langle \cos(n \phi)\rangle$ is exactly $0$ at $q_\perp = 0$, and at small $q_\perp$ they scale as $\langle \cos(n \phi)\rangle \propto q_\perp^n$ \cite{Catani:2017tuc}. 
At large $q_\perp$, where the hard gluon radiation dominates and there is no preferred direction, $\langle \cos(n \phi)\rangle$ must decrease as functions of $q_\perp$.
This will modify the behavior of both $\langle \cos(\phi)\rangle$ and $\langle \cos(2\phi)\rangle$ asymmetries at large $q_\perp$. 
Besides this, the $\langle \cos(\phi)\rangle$ and $\langle \cos(2\phi)\rangle$ predictions present a clear difference.
The former highly depends on the model considered, with the NRQCD being more suppressed compared to the CSM. 
We see that the NRQCD suppression holds also upon inclusion of the LDME evolution, but in addition we observed that certain combinations of LDMEs can also lead to a change of sign compared to the CSM.
On the other hand, the CSM and NRQCD outcomes are comparable, and in general non-negligible, for $\langle \cos(2\phi)\rangle$.

As demonstrated, such differences offer a unique opportunity to determine the underlying quarkonium production mechanism.
In particular, a suppressed $\langle \cos(\phi)\rangle$ with a sizable $\langle \cos(2\phi)\rangle$ implies that contributions driven by the CO channel are important at low-$q_\perp$ and one must then include them, e.g.,~by means of the NRQCD factorization. Once asserted, data relative to these observables can be incorporated in global fits to further constrain LDME values. In addition, one can also verify the consistencies with different observables, which might be an indication of the presence of factorization-breaking effects.

\section{Conclusions} 
\label{sec: conclusions}

In summary, we have demonstrated that the soft gluon radiation leads to significantly different azimuthal angular correlations between the CS and CO mechanisms in the $J/\psi$ plus jet photoproduction process at the future EIC. We have shown that such differences hold for both gluon and quark channels. This directly affects azimuthal asymmetry predictions within the CSM and the NRQCD approach, where the latter presents a soft dependence on the LDME. We also expect that including higher Fock-states' contributions will not modify the above conclusion.
Thus, we consider these observables as the stems to disentangle these two production mechanisms and investigate the absence of NRQCD factorization-breaking effects at small $q_\perp$.
We also expect that similar conclusions can be drawn for other experiments as well, like in hadronic processes at the LHC.
Moreover, $\cos(2\phi)$ has been proposed to study the linearly polarized gluon distribution in the electroproduction of $J/\psi$ plus jet at the EIC. We expect that the soft gluon radiation will lead to sizable contributions to $\cos(2\phi)$ asymmetry in this process as well, with a behavior similar to that shown in Fig.~\ref{fig: asymmetries NRQCD} for photoproduction. 
Understanding the soft gluon radiation will be a crucial step to unambiguously determine the gluon tomography of linearly polarized distribution from this measurement. 
We will explore all these physics, including higher-order perturbative and power corrections, in the future.

\section*{Acknowledgments}  
We thank Prof. D. Boer and Dr. M. Nefedov for the valuable discussions.
This material is based upon work supported by the U.S. Department of Energy, Office of Science, Office of Nuclear Physics, under contract numbers DE-AC02-05CH11231.

\bibliography{reference.bib}

\clearpage
\newpage

\appendix 

\section{Azimuthal asymmetries in the CS and CO channels separately}
\label{app: CS vs CO}
In this appendix we present direct comparisons between the CS and CO mechanisms, providing more insights on the results shown in Fig.~\ref{fig: asymmetries NRQCD}.

\begin{figure}[t]
\vspace{1em}
\centering
\includegraphics[width=\linewidth]{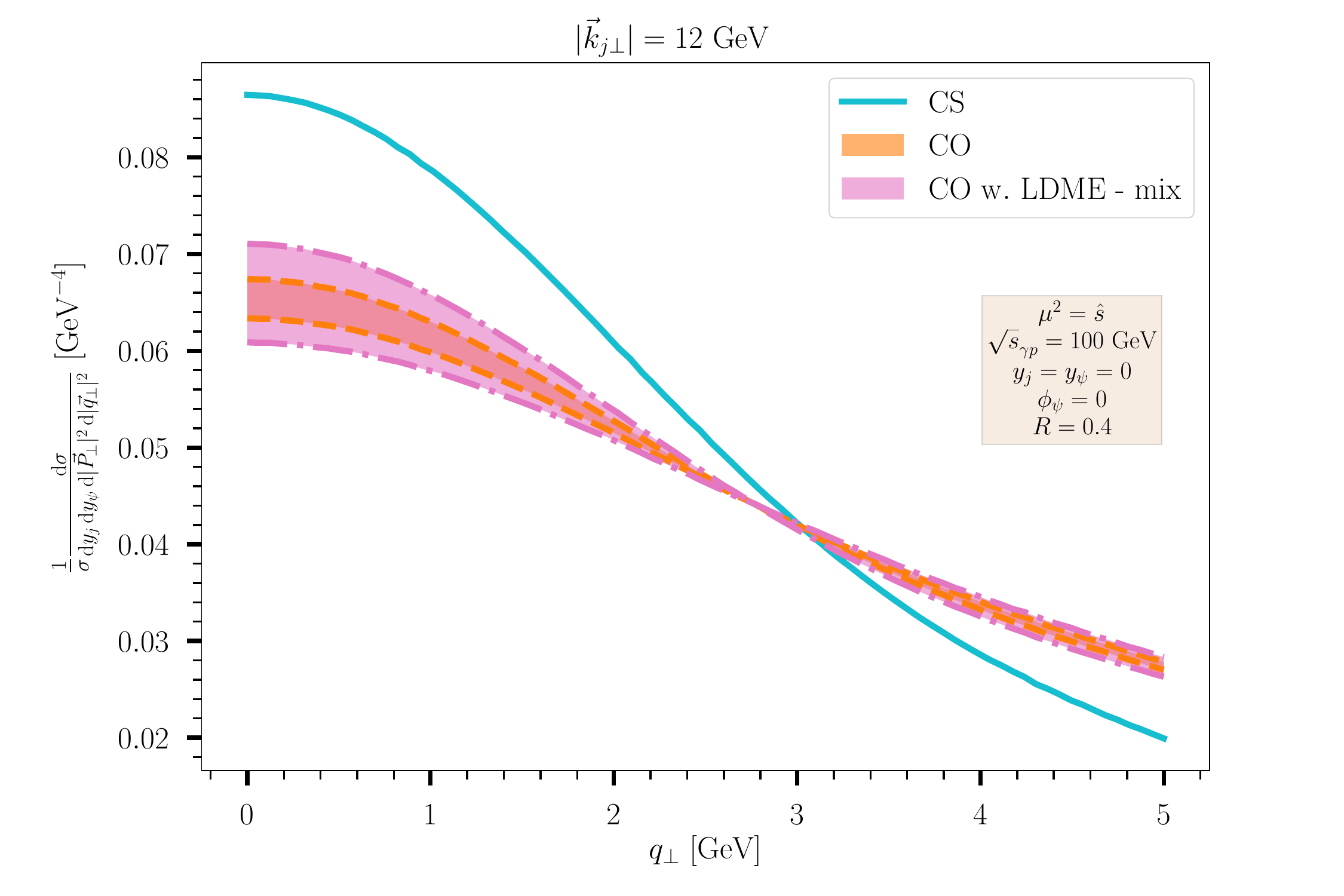}
\vskip -0.2cm \caption{\it Normalized differential cross section of isotropic $J/\psi$ plus jet photoproduction at $\sqrt s_{\gamma p} = 100$~GeV in a frame where the azimuthal angle of $\vec P_\perp$ is zero and $y_j = y_\psi = 0$. We considered two values of $|\vec k_{j\perp}|$, while $R = 0.4$. Solid cyan line corresponds to the CS channel. Orange and pink bands correspond to the CO one, with the latter including the contribution from the LDME evolution.}
\label{fig: production CS vs CO}
\end{figure} 

In addition to the asymmetries, in Fig.~\ref{fig: production CS vs CO} we present the normalized differential cross section within both the CS and CO channels and for the same kinematics considered for Fig.~\ref{fig: asymmetries NRQCD}.
The CO channel has a much wider distribution as compared to the CS one as a result of the final state gluon radiation associated with the heavy quark pair, which leads to a significant difference in the associated $C_0^{(c)}$ coefficients (see Table~\ref{tab: coefficients}). 
The bands in the CO channel are constructed by combining the central values of several LDME sets (see also Fig.~\ref{fig: asymmetries NRQCD}). As expected, this uncertainty increases once the contribution of the LDME evolution is taken into account. We point out that for the kinematics discussed here such contribution mostly arises from the quark channel, while we have found it negligible for the gluon channel.
Moreover, it is worth mentioning that Fig.~\ref{fig: production CS vs CO} is obtained for $y_j = y_\psi = 0$ and different rapidities values can slightly modify the picture, e.g. less broadened distributions for $y_j = 1$. Nonetheless, other choices of $y_j$ and $y_\psi$ do not spoil the main conclusion of Fig.~\ref{fig: production CS vs CO}, namely that we identify different shapes of the normalized differential cross sections within the CS and CO mechanisms.

\begin{figure}[t]
\begin{center}
\subfloat[\label{fig: asy PT12}]{\includegraphics[width=.95\linewidth, keepaspectratio]{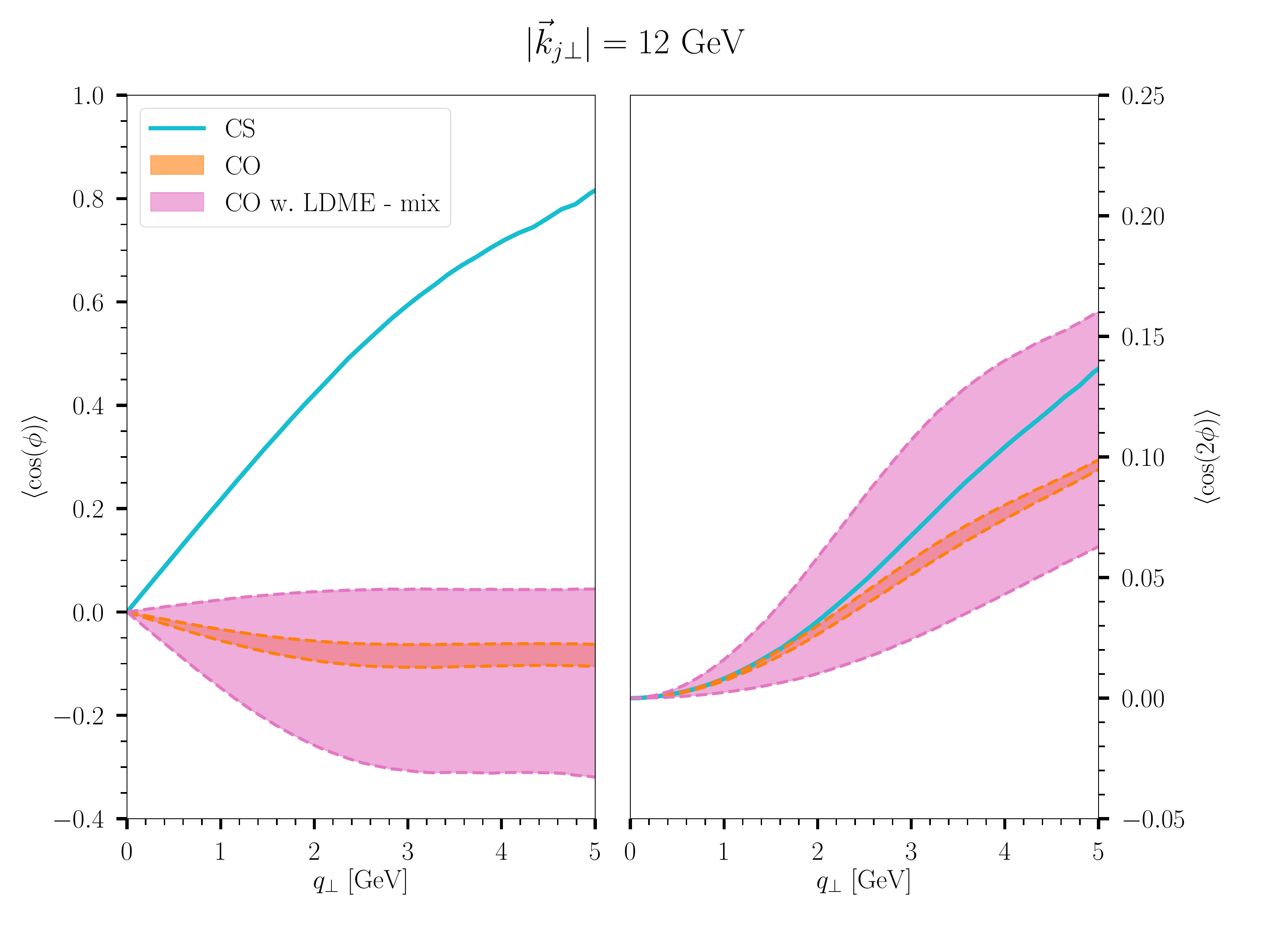}}
\end{center}
\vskip -0.4cm \caption{\it Averaged azimuthal asymmetries for $J/\psi$ plus jet photoproduction at $\sqrt s_{\gamma p} = 100$~GeV and $k_{j\perp}| = 12$~GeV, obtained within the CS (solid cyan line) and CO (orange and pink bands) mechanisms. Jet size is $R = 0.4$.}
\label{fig: asymmetries CS vs CO}
\end{figure}

Moving to the asymmetries, in Fig.~\ref{fig: asymmetries CS vs CO} we present $\langle \cos(\phi)\rangle$ and $\langle \cos(2\phi)\rangle$ predictions in the CS and CO channels as functions of $q_\perp$.
We consider $k_{j\perp} = 12~{\rm GeV}$ as done in Fig.~\ref{fig: asymmetries NRQCD}.
While the CS channel is exclusively driven by gluons, in the CO one we have the interplay between quarks and gluons. If one does not include the LDME mixing, the quark contribution (at $k_{j\perp} = 12~{\rm GeV}$) is negligible, causing a suppression of the dependence on the LDME choice. Consequently, the variation observed in the full NRQCD result is mostly due to the significance of the CO mechanism with respect to the CS one. 
On the other hand, upon the inclusion of the LDME evolution, the quark channel becomes more significant, and the uncertainty band driven by the choice of the LDME set opens up. Not that the magnitude of the $\langle \cos(\phi)\rangle$ asymmetry in the CO channel is always lower than the CS case, while a change of sign (especially upon the inclusion of the LDME evolution) might be present.
At variance,
$\langle \cos(\phi)\rangle$ predictions employing the two mechanisms are always comparable to each other, and they undoubtedly agree in sign.

\end{document}